\documentclass[10pt,twocolumn,landscape,a4paper]{article}

\font\mybb=msbm10 at 10pt
\def\bb#1{\hbox{\mybb#1}}
\def\Z {\bb{Z}}
\def\R {\bb{R}}
\begin{document}
\begin{titlepage}

\begin{flushright}
CERN-TH/95--348\\
McGill/95-62\\
{\bf hep-th/9512177}\\
December $21$st, $1995$\\
(Revised February $2$nd $1996$)
\end{flushright}

\begin{center}

%title
%%%%%%%%%%%%%%%%%%%%%%%%%%%%%%%%%%%%%%%%%%%%%%%%%%%%%%%%%%%%%%%%%%%%%%%%
%%% DO NOT CHANGE THIS ONE (SEE BELOW)
%%%%%%%%%%%%%%%%%%%%%%%%%%%%%%%%%%%%%%%%%%%%%%%%%%%%%%%%%%%%%%%%%%%%%%%%
\baselineskip25pt
%%%%%%%%%%%%%%%%%%%%%%%%%%%%%%%%%%%%%%%%%%%%%%%%%%%%%%%%%%%%%%%%%%%%%%%%
{\LARGE {\bf SUPERSYMMETRIC BLACK HOLES IN $N=8$ SUPERGRAVITY}}

\vspace{1cm}

%authors
{\large{\bf Ramzi R.~Khuri}${}^{a,b}$
\footnote{E-mail address: {\tt khuri@nxth04.cern.ch}}
        {\bf and Tom\'as Ort\'{\i}n}${}^{a}$
\footnote{E-mail address: {\tt tomas@surya20.cern.ch}}\\

\vspace{.5cm}

${}^{a}${\it C.E.R.N.~Theory Division}\\
{\it CH--1211, Gen\`eve 23, Switzerland}\\

\vspace{.5cm}

${}^{b}${\it McGill University}\\
{\it Physics Department}\\
{\it Montreal, PQ, H3A 2T8 Canada}\\
}
\end{center}

\vspace{1.5cm}

%%%%%%%%%%%%%%%%%%%%%%%%%%%%%%%%%%%%%%%%%%%%%%%%%%%%%%%%%%%%%%%%%%%%%%

\begin{abstract}

We study the embedding of extreme (multi-) dilaton black hole solutions
for the values of the parameter $a=\sqrt{3},1,1/\sqrt{3},0$ in $N=4$ and
$N=8$ four-dimensional supergravity.  For each black hole solution we
find different embeddings in $N=4$ supergravity which have different
numbers of unbroken supersymmetries.  When embedded in $N=8$
supergravity, all different embeddings of the same solution have the
same number of unbroken supersymmetries.  Thus, there is a relation
between the value of the parameter $a$ and the number of unbroken
supersymmetries in $N=8$ supergravity, but not in $N=4$, and the
different embeddings must be related by dualities of the $N=8$ theory
which are not dualities of the $N=4$ theory.  The only exception in this
scheme is a {\it dyonic} embedding of the $a=0$ black-hole solution
which seems to break all supersymmetries both in the $N=4$ and in the
$N=8$ theories.

\end{abstract}

\vspace{1cm}

\begin{flushleft}
CERN-TH/95--348\\
McGill/95-62\\
\end{flushleft}

\end{titlepage}

\newpage

%%%%%%%%%%%%%%%%%%%%%%%%%%%%%%%%%%%%%%%%%%%%%%%%%%%%%%%%%%%%%%%%%%%%%%%%
%%% CHANGE THIS ONE
%%%%%%%%%%%%%%%%%%%%%%%%%%%%%%%%%%%%%%%%%%%%%%%%%%%%%%%%%%%%%%%%%%%%%%%%
\baselineskip11pt
%%%%%%%%%%%%%%%%%%%%%%%%%%%%%%%%%%%%%%%%%%%%%%%%%%%%%%%%%%%%%%%%%%%%%%%%
\pagestyle{plain}

%%%%%%%%%%%%%%%%%%%%%%%%%%%%%%%%%%%%%%%%%%%%%%%%%%%%%%%%%%%%%%%%%%%%%%

\section{Introduction}

The abundance of extreme black-hole and soliton solutions of string
theory is intimately connected with the existence of duality symmetries
either within a given string theory or between different string and
higher-membrane theories (see \cite{kn:DK,kn:PR,kn:HT,kn:W} and
references therein).  These duality symmetries can always be used as
solution-generating transformations which generate new solutions out of
known solutions, and, therefore, the larger the duality group, the
greater the number of extreme black-hole and soliton solutions.

There is, however, another point of view, which may be regarded as
``dual'' to this one: solutions related by duality are in some sense
equivalent (some times they are equivalent in a strict sense), so that
solutions generated by duality transformations are not regarded as
completely new.  From this latter viewpoint, one is particularly
interested in solutions which are not related by any duality symmetry
since new solutions can then be generated from this reduced set of
inequivalent solutions.  In this scenario, the larger the group of
duality symmetries, the smaller the set of inequivalent solutions.

This is the point of view that we are going to adopt in this paper.  We
consider extreme black-hole solutions arising in four-dimensional $N=4$
and $N=8$ supergravity reductions of $N=1$ and $N=2$ ten-dimensional
supergravities.  Our main goal is the following: we want to find how
many inequivalent extreme dilaton black-hole solutions there are in
these theories, with just one scalar and one vector field in four
dimensions and coupling between the scalar and the vector field
characterized by the constant $a$.  Equivalent solutions should be
related by duality symmetries and should have the same number of
unbroken supersymmetries.

In fact, the interplay between duality and supersymmetry, both spacetime
and worldsheet, has been the subject of active investigation recently.
In particular, it has been noted that soliton solutions of a given
theory which transform into each other under $T$~duality preserve the
same amount of supersymmetry under most circumstances
\cite{kn:BBKOBSHAAGB}. $N=4,D=4$ Killing spinors transform covariantly
under $S$~duality, the number of unbroken supersymmetries and
Bogomol'nyi bounds being invariant \cite{kn:O1,kn:O2}, and similar
results have been obtained in the context of $N=4,D=8$ supergravity
\cite{kn:IPT}.  More recently, an analysis of the supersymmetry-breaking
pattern of string-like solitons in toroidally compactified
four-dimensional heterotic string theory \cite{kn:SS} showed that such
solitons which are related by an $O(8,24;\Z)$ transformation (this
larger duality group containing both the $SL(2,\Z)$ $S$~duality group
and the $O(6,22;\Z)$ $T$~duality group) preserved the same amount of
spacetime supersymmetry, whether it was $1/2$, $1/4$ or $1/8$ of the
original supersymmetries of the theory.

Thus, supersymmetry can be used to classify inequivalent extreme
black-hole solutions.

Most of the extreme black-hole solutions discussed so far have been
found to preserve precisely half of the supersymmetries of whatever
low-energy theory they arise in: extreme Reissner-Nordstr\"om black
holes break one half of the $N=2,D=4$ supersymmetries (see, for
instance, the second lecture of Ref.~\cite{kn:GWG} and references
therein), ($a=1$) extreme dilaton black holes break one half of the
$N=4,D=4$ supersymmetries \cite{kn:KLOPP}, Kaluza-Klein black holes
break one half of the supersymmetries of $N=1,D=5$ supergravity since
they correspond to five-dimensional $pp$-waves (see, for instance, the
third lecture of Ref.~\cite{kn:GWG} and references therein).
Nevertheless, there exist several examples of solutions which break
$3/4$ of the supersymmetries or more, such as the double-instanton
string of \cite{kn:DI} and most of the string-like solitons of
\cite{kn:SS} (see also \cite{kn:PR} and references therein for other
examples)

However, the statement that the ($a=1$) extreme dilaton black hole has
one half of the $N=4,D=4$ supersymmetries unbroken (that is half of
$N=1,D=10$) is not, in fact, true, or, more precisely, is incomplete.
One should say that there is an {\it embedding} of the ($a=1$) extreme
dilaton black hole which has one half of the $N=4,D=4$ supersymmetries
unbroken.  This is the embedding proposed in Ref.~\cite{kn:KLOPP} and
basically corresponds to the identification of the vector field with a
vector field belonging to the gravity supermultiplet of $N=4,D=4$
supergravity.  There is, however, another embedding of the same solution
which is {\it not} supersymmetric, and which corresponds to the
identification of the vector field with one belonging to an $N=4,D=4$
vector supermultiplet\footnote{This vector supermultiplet can have a
Kaluza-Klein origin, appearing in the dimensional reduction of
$N=1,D=10$ supergravity or can be related to one of the sixteen
ten-dimensional $U(1)$ vector supermultiplets of the heterotic string.}
\cite{kn:GHS}.

These two embeddings are, according to the above results on duality and
supersymmetry, inequivalent in the framework of the $N=4$ theory, in
spite of the fact that they correspond to identical solutions satisfying
identical Bogomol'nyi-like bounds.

This situation seems to us a bit paradoxical.  One would naively think
that all the possible embeddings of a given solution should be
equivalent.  One would also naively think that all Bogomol'nyi-like
bounds should be related to supersymmetry, at least in some theory.

As we will see, the resolution of this apparent paradox lies in the
$N=8$ theory.  All embeddings in the $N=4$ theory can also be considered
as embeddings in the $N=8$ theory, and we will only consider this kind
of simultaneous (or NS-NS) embedding.  With the exception of a new
dyonic black hole which breaks all of the supersymmetries \cite{kn:KO},
our findings indicate that there is a well-defined relation between the
parameter $a$ and the number of supersymmetries preserved in $N=8$
supergravity.  For either purely electric or purely magnetic
four-dimensional configurations, all NS-NS embeddings of the same
extreme black-hole solution have the same number of unbroken
supersymmetries and this suggests that all embeddings are equivalent in
the framework of the $N=8$ theory.  Some of the $N=8$ dualities
($U$~dualities \cite{kn:HT}) that connect the different embeddings are
not present in the $N=4$ theory, and so it follows that these embeddings
are not $(N=4)$-equivalent.

Particularly interesting is the $U$~duality transformation that we call
$C$~duality whose effect is to interchange supergravity and matter
vector fields and the chirality of the ten-dimensional spinors
\cite{kn:BJO}.  The embeddings proposed in Refs.~\cite{kn:GHS,kn:KLOPP}
are related by $C$~duality and therefore are both supersymmetric in the
$N=8$ theory.  However, $N=1,D=10$ supergravity is chiral, and
$C$~duality cannot be a duality symmetry of the $N=4$ theory.  In fact,
$C$~duality is a string/string duality relating the two $N=1$
supergravity theories with opposite chiralities that one can construct
in ten dimensions.  This explains the apparent paradox.

Four-dimensional Kaluza-Klein black holes have been studied in 
Refs.~\cite{kn:CYKK}, while four-dimensional black-hole solutions of $N=8$
supergravity have also been studied in Refs.~\cite{kn:CYN8}. Furthermore, 
in Refs.~\cite{kn:CY4P}, a class of regular BPS saturated black holes
parameterized by four charge parameters was constructed, while in 
Refs.~\cite{kn:CY5P} a five-parameter construction of all static,
spherically symmetric BPS-saturated black holes of heterotic string theory
compactified on a six-torus was shown. While our solutions appear to be 
special examples of the abovementioned generalized constructions, 
our work differs from these in that
we are interested in the interplay between $U$~duality and the different
truncations of $N=8$ supergravity. The appearance of supersymmetric
extreme solutions which saturate a Bogomol'nyi bound and non-supersymmetric 
extreme solutions which do not saturate a Bogomol'nyi bound in the context 
of the four-parameter solution of Refs.~\cite{kn:CY4P} was shown in 
Refs.~\cite{kn:CYSS} to arise from the different limits from non-extremality
of the four-parameter solution. This observation does not, however, resolve
the apparent paradox in question in this paper, as all the classes of black
hole solutions we consider do in fact saturate a Bogomol'nyi-like bound.

A summary of our work is as follows: in Section~\ref{sec-amodel} we
write down the action and multi-black hole solutions of the $a$-model,
noting for which values of $a$ the solutions arise from consistent
truncations of maximal $N=8$ supergravity.  In
Section~\ref{sec-embedding} we demonstrate the embeddings, both known
and novel, of the $a$-model solutions in $N=4$ and $N=8$ supergravity,
and proceed in Section~\ref{sec-supersymmetry} to find the unbroken
supersymmetries for each embedding.  Finally, we discuss our results in
Section~\ref{sec-conclusion}.

The appendices contain some complementary material and results that we
heavily use in the main body of the paper: our conventions are in
Appendix~\ref{sec-conventions}, in Appendix~\ref{sec-IIAsusy} we derive
from eleven dimensions the ten-dimensional supersymmetry rules of the
type~IIA theory and of the two $N=1$ theories of opposite chiralities,
together with the $C$~duality relation between them.
Appendix~\ref{sec-spincon} contains the spin connection one-form for the
ten-dimensional class of metrics which we consider.

%%%%%%%%%%%%%%%%%%%%%%%%%%%%%%%%%%%%%%%%%%%%%%%%%%%%%%%%%%%%%%%%%%%%%%

\section{The ``$a$-model''}
\label{sec-amodel}

In reducing to four dimensions the ten-dimensional $N=1$ and $N=2$
supergravities arising in the low-energy limit of the various
superstring theories, one generically arrives at a complicated
four-dimensional action with many scalar fields (moduli) and Maxwell
vector fields (throughout this work we consider Abelian vector fields
only, as even when we start with a non-Abelian gauge group, a generic
point in the moduli space has $U(1)^n$ symmetry).  For example,
toroidally compactified four-dimensional heterotic string theory
consists of a metric, antisymmetric tensor, dilaton, $28$ vectors and
$132$ scalars.

In this paper we consider a greatly simplified truncation in
four-dimensions, consisting of a metric
$\tilde{g}_{\mu\nu}$, a single scalar field $\varphi$ and a single
Maxwell field $A_{\mu}$ with an arbitrary parameter $a$ governing the
scalar-Maxwell coupling.  The ``$a$-model'' action is explicitly given
by

\begin{equation}
S^{(a)}= {\textstyle\frac{1}{2}}\int d^{4}x\ \sqrt{|\tilde{g}|}
\left[-\tilde{R} -2(\partial\varphi)^{2}
+{\textstyle\frac{1}{2}}e^{-2a\varphi}F^{2}\right]\, .
\end{equation}

We stress that the scalar $\varphi$ is in general different from the
string dilaton, and is in fact a linear combination of the dilaton and
other moduli.  Throughout we always denote the dilaton by a different
symbol ($\phi$).  Therefore, in working with this model, we are always
in the canonical (Einstein) frame, which we denote with tildes on the
metric, Einstein tensor etc.  The equations of motion are

\begin{equation}
\left.
\begin{array}{rcl}
\tilde{G}_{\alpha\beta} +2 T_{\alpha\beta}^{\varphi} -e^{-2a\varphi}
T_{\alpha\beta}^{A} & = & 0\, ,
\\
& &
\\
\tilde{\nabla}^{2}\varphi -{\textstyle\frac{a}{4}}e^{-2a\varphi}F^{2} &
= & 0\, ,
\\
& &
\\
\tilde{\nabla}_{\mu}\left(e^{-2a\varphi} F^{\mu\alpha}  \right) & = &
0\, ,
\end{array}
\right\}
\label{eq:aeqmo}
\end{equation}

\noindent where $T^{\varphi}_{\alpha\beta}$ and $T^{A}_{\alpha\beta}$
are the energy-momentum tensors of the scalar field $\varphi$ and the
vector field $A_{\mu}$ respectively:

\begin{eqnarray}
T^{\varphi}_{\alpha\beta} & = & \partial_{\alpha}\varphi
\partial_{\beta}\varphi -{\textstyle\frac{1}{2}}\tilde{g}_{\alpha\beta}
(\partial\varphi)^{2}\, ,
\label{eq:tscalar} \\
& &
\nonumber \\
T^{A}_{\alpha\beta} & = &F_{\alpha}{}^{\mu} F_{\beta\mu}
-{\textstyle\frac{1}{4}}\tilde{g}_{\alpha\beta}F^{2}\, .
\label{eq:tvector}
\end{eqnarray}

Black-hole solutions of the $a$-model exist for all values of the
parameter $a$, that we take to be positive without any loss of
generality.  In particular, there are extreme \cite{kn:HW} and
multi-black-hole solutions \cite{kn:S1,kn:O1} for all $a$.  The purely
electric extreme multi-black-hole solutions are

\begin{equation}
\left\{
\begin{array}{rcl}
d\tilde{s}^{2} & = & V^{-\frac{2}{1+a^{2}}} dt^{2}
-V^{+\frac{2}{1+a^{2}}}d\vec{x}^{2}\, ,
\\
& &
\\
e^{\varphi} & = & V^{-\frac{a}{1+a^{2}}}\, ,
\\
& &
\\
F_{t\underline{i}} & = & -n\ \sqrt{\frac{2}{1+a^{2}}}\
\partial_{\underline{i}}V^{-1}\, ,
\end{array}
\right.
\label{eq:mbhasolutions}
\end{equation}

\noindent where $V(\vec{x})$ is a harmonic function in three-dimensional
Euclidean space

\begin{equation}
\partial_{\underline{i}} \partial_{\underline{i}}V=0\, ,
\end{equation}

\noindent that we always take to be positive and normalized so as
to make the above metric regular and asymptotically flat, that is

\begin{equation}
V(\vec{x})= 1 +\sum_{i}\frac{M_{i}}{|\vec{x}-\vec{x}_{i}|}\, ,
\hspace{.5cm}M_{i}\geq 0\hspace{.2cm}\forall i\, ,
\end{equation}

\noindent and $n=\pm 1$ gives the sign of the electric charges.
The equations of motion of the $a$-model are invariant under the
discrete electric-magnetic duality transformation

\begin{equation}
F^{\prime} =  e^{-2a\varphi} F\, ,
\hspace{1cm}
\varphi^{\prime}  =  -\varphi\, ,
\label{eq:amemdual}
\end{equation}

\noindent and, therefore, a purely magnetic multi-black-hole solution
always exists for any $a$:

\begin{equation}
\left\{
\begin{array}{rcl}
\tilde{ds}^{2} & = & W^{-\frac{2}{1+a^{2}}} dt^{2}
-W^{+\frac{2}{1+a^{2}}}d\vec{x}^{2}\, ,
\\
& &
\\
e^{\varphi} & = & W^{+\frac{a}{1+a^{2}}}\, ,
\\
& &
\\
F_{\underline{i}\underline{j}} & = & \mp\sqrt{\frac{2}{1+a^{2}}}\
\epsilon_{ijk}\partial_{\underline{k}}W\, .
\end{array}
\right.
\end{equation}

For the special values $a=0$ and $a=1$ dyonic solutions also exist
\cite{kn:GGM,kn:KLOPP}:

\begin{equation}
\left\{
\begin{array}{rcl}
\tilde{ds}^{2} & = & (VW)^{-1} dt^{2}-VWd\vec{x}^{2}\, ,
\\
& &
\\
e^{\varphi} & = & V^{-\frac{1}{2}}W^{+\frac{1}{2}}\, ,
\\
& &
\\
F & = & n\ dV^{-1}\wedge dt
-{\textstyle\frac{1}{2}}m\ \epsilon_{ijk} \partial_{\underline{i}}W
dx^{\underline{j}}\wedge dx^{\underline{k}}\, ,
\end{array}
\right.
\end{equation}

\noindent where $n$ and $m$ take the values $\pm 1$. (The $a=0$ dyon
is obtained by setting $V=W$ in the above solution.)

All the purely electric or magnetic extreme solutions (and the dyonic
$a=1$, $a=0$ solutions) admit Killing spinors if one uses the
appropriate definition of ``gravitino'' and ``dilatino'' supersymmetry
transformation rules
 \cite{kn:GKLTT}. These rules coincide with the supersymmetry rules of
known supergravity theories in some cases, and they can always be used
to do Nester constructions. It is worth noting, though,
that all the supersymmetry rules of these subsupergravities can be
obtained from the $N=4,D=4$ supersymmetry rules (with no axion)
through the same transformation that takes the $a=1$ extreme dilaton
black hole into the other values of $a$ (see the conclusions
section of Ref.~\cite{kn:O1}).

However, it is not known which values of $a$ naturally appear in {\it
true} supergravity theories.  As explained in Ref.~\cite{kn:HT}, some of
them are expected to arise in the different consistent truncations of
maximal $N=8$ supergravity, namely those with $a=\sqrt{3}, 1,
1/\sqrt{3}, 0$.  The values $1$ and $0$ arise in the truncations to
$N=4$ and $N=2$ respectively.  The values $\sqrt{3}$ \cite{kn:HM} and
$1/\sqrt{3}$ arise in the truncation from maximal to simple
five-dimensional supergravity and its dimensional reducation to $D=4$.
For all four values of $a$ one can also argue that the extreme black-hole
solutions are solitonic.

A study of the slow motion of the extreme black-hole solutions of the
$a$-model reveals that only for $a=\sqrt{3}$ the corresponding moduli
space is flat \cite{kn:K,kn:S2}.  This is the necessary condition for
the solutions to only break half of the $N=8$ supersymmetries \cite{kn:G}
({\it i.e.} to be BPS states), and therefore we only expect these to
have half of the $N=8$ supersymmetries unbroken and the rest will have
fewer unbroken supersymmetries.

It was conjectured in \cite{kn:DR} that certain electrically charged
extreme black holes arising in the $N=4$ theory could be identified with
BPS states in the spectrum of allowed charges of the theory
\cite{kn:SCH} (the so-called Schwarz-Sen spectrum), which in turn could
be identified with elementary (massive) string states.  In the $N=4$
theory, both the $a=\sqrt{3}$ black hole and a certain embedding of the
$a=1$ black hole preserve half of the spacetime supersymmetries, and
were shown to correspond to massive string states (dynamical evidence
supporting the conjecture in \cite{kn:DR} was found in \cite{kn:KM}).

In the $N=8$ theory, however, only $a=\sqrt{3}$ black holes preserve
half the supersymmetries, and are therefore the only candidates to be
identified with elementary string states.  On the other hand, all four
black holes (at least in some embeddings) preserve some degree of
supersymmetry, and saturate Bogomol`nyi bound formulae (see, for
example, \cite{kn:CY4P,kn:LIU}).  In truncating to an $N=2$ theory, one 
can find
embeddings for all four black holes such that each preserves half of the
spacetime supersymmetries.  As a result, there is the possibilty of
realizing all these extremal black holes as string states, although in
the $N=2$ case the solutions are no longer protected by
nonrenormalization theorems against quantum corrections.

The problem with quantum corrections arises especially for the $a=0$
black hole, since this solution has zero dilaton for both electric and
magnetic solutions, and there is no way to distinguish a perturbative
from a non-perturbative state.  One is then led to conclude that both
solutions are non-perturbative, and cannot correspond to an elementary
string state to begin with\footnote{This was pointed out to us by Paul
Townsend.}.

On the other hand, it was also conjectured in Ref.~\cite{kn:DR} that the
$a=1,1/\sqrt{3},0$ extreme dilaton black holes could be understood as
bound states of the $2, 3$ and $4$ maximally supersymmetric $a=\sqrt{3}$
black holes respectively.  This conjecture has been recently confirmed
in Ref.~\cite{kn:R} where it was shown how to get the $a=1,1/\sqrt{3},0$
solutions as particular limits of a multi-$a=\sqrt{3}$-black-hole
solution that interpolates between them.  In our supersymmetry analysis,
a similar picture of compositeness arises in relating the Killing spinor
equations of the various black holes.  For example, as we shall see
below, the supersymmetry breaking pattern of the $a=1/\sqrt{3}$ black
hole, corresponding to $3$ $a=\sqrt{3}$ black holes in the bound state
picture, arises as the combination of the supersymmetry breaking of a
single $a=\sqrt{3}$ black hole and an $a=1$ black hole, corresponding to
$2$ $a=\sqrt{3}$ black holes in the bound state picture.

Additional information on the $a$-model comes from its reduction to two
dimensions. For (and only for) the special values $a=0,\sqrt{3}$ the two
dimensional theory has infinite symmetry and becomes completely
integrable \cite{kn:GGK}.

Our purpose in the next section is to investigate which solutions of the
$a$-model do arise in $N=4(8)$, and for which values of $a$, how they
are embedded in the supergravity theory and their unbroken
supersymmetries.

%%%%%%%%%%%%%%%%%%%%%%%%%%%%%%%%%%%%%%%%%%%%%%%%%%%%%%%%%%%%%%%%%%%%%%

\section{Embedding the $a$-model solutions in $N=4(8)$ supergravity}
\label{sec-embedding}

We are ultimately interested in the embedding of the extreme (multi-)
black-hole solutions of the $a$-model into $N=8,D=4$ supergravity, which
is equivalent (upon dimensional reduction) to any of the $N=2,D=10$
supergravities.  For simplicity, we will focus on the
Neveu-Schwarz-Neveu-Schwarz (NS-NS) subsector of this theory, which can
be obtained by simply setting to zero all the Ramond-Ramond (R-R)
fields.  As explained in Ref.~\cite{kn:BHO} and in
Appendix~\ref{sec-IIAsusy}, where more details can be found with the
explicit example of the type~IIA theory, this truncation is consistent
({\it i.~e.}~any solution of the truncated theory is automatically a
solution of the original one) and the bosonic sector of the truncated
theory is the bosonic sector of the $N=1,D=10$ supergravity theory.

The consistency of the truncation has a stringy explanation: the only
sources for R-R fields have to be R-R fields. Therefore, there are no
purely NS-NS terms in the equations of motion of the R-R fields and all
terms simultaneously vanish, leaving no constraints.

Dimensional reduction of $N=1,D=10$ supergravity to $D=4$ gives
$N=4,D=4$ supergravity coupled to six matter (vector) supermultiplets
\cite{kn:C}.  Therefore, solutions of $N=4,D=4$ supergravity coupled to
six vector supermultiplets can be considered simultaneously as solutions
of $N=8,D=4$ (or $N=2,D=10$) supergravity.  Since the supersymmetry
transformation rules are much simpler in ten dimensions, we will uplift
any solution of the $N=4 (+6V),D=4$ theory to ten dimensions to get
solutions of the $N=1,2$ theories.

In this section we want to identify further {\it consistent} truncations
of the $N=4 (+6V),D=4$ theory that lead us to the $a$-model for some
values of $a$, so, in the end, and following the logic of the previous
paragraph, we have a solution of the $N=1(2),D=10$ theory for each
solution of the $a$-model.

To study the consistency of the truncations, we need the equations of
motion.  They could be derived from the action Eq.~(\ref{eq:4daction}).
However, since the $a$-model makes sense only in the canonical metric,
we first rescale the string metric $g_{\mu\nu}$ in
Eq.~(\ref{eq:4daction}) to the canonical metric $\tilde{g}_{\mu\nu}=
e^{-2\phi} g_{\mu\nu}$ and get

\begin{eqnarray}
\tilde{G}_{\alpha\beta} +2 T_{\alpha\beta}^{\phi}
+{\textstyle\frac{9}{4}} T_{\alpha\beta}^{B}
& &
\nonumber \\
& &
\nonumber \\
-{\textstyle\frac{1}{4}} \left[\partial_{\alpha}G_{mn}
\partial_{\beta} G^{mn} -{\textstyle\frac{1}{2}}
\tilde{g}_{\alpha\beta} \partial_{\mu}G_{mn} \partial^{\mu}G^{mn}
\right]
& &
\nonumber \\
& &
\nonumber \\
-{\textstyle\frac{1}{4}} G^{mn} G^{pq} \left[\partial_{\alpha}B_{mp}
\partial_{\beta} B_{nq} -{\textstyle\frac{1}{2}}
\tilde{g}_{\alpha\beta} \partial_{\mu}B_{mp} \partial^{\mu}B_{nq}
\right]
& &
\nonumber \\
& &
\nonumber \\
+{\textstyle\frac{1}{2}} G_{mn} \left[F^{(1)m}{}_{\alpha}{}^{\mu}
F^{(1)m}{}_{\beta\mu} -{\textstyle\frac{1}{4}}
\tilde{g}_{\alpha\beta}F^{(1)m}{}_{\mu\nu} F^{(1)m\mu\nu} \right]
& &
\nonumber \\
& &
\nonumber \\
+{\textstyle\frac{1}{2}} G^{mn} \left[{\cal F}_{m\alpha}{}^{\mu}
{\cal F}_{n\beta\mu} -{\textstyle\frac{1}{4}}\tilde{g}_{\alpha\beta}
{\cal F}_{m\mu\nu} {\cal F}_{n}{}^{\mu\nu} \right]
& = & 0\, ,
\label{eq:n4d4eqmofirst} \\
& &
\nonumber \\
\tilde{\nabla}^{2}\phi +{\textstyle\frac{3}{4}}e^{-4\phi}H^{2}
+{\textstyle\frac{1}{8}}e^{-2\phi}\left[G_{mn}F^{(1)m} F^{(1)n}
+G^{mn}{\cal F}_{m} {\cal F}_{n}\right]
& = & 0\, ,
\\
& &
\nonumber \\
\tilde{\nabla}^{2}G^{rs} -G^{m(r}G^{s)n}G^{pq}\left[\partial G_{mp}
\partial G_{nq} +\partial B_{mp} \partial B_{nq} \right]
& &
\nonumber \\
& &
\nonumber \\
+{\textstyle\frac{1}{2}} e^{-2\phi}\left[F^{(1)r}F^{(1)s}
-G^{m(r}G^{s)n}{\cal F}_{m} {\cal F}_{n}\right]
& = & 0\, ,
\\
& &
\nonumber \\
\hat{\nabla}_{\mu}\left(G^{nr}G^{qs}\partial^{\mu}B_{nq} \right)
+e^{-2\phi}{\cal F}_{m}G^{m[s}F^{(1)r]}
& = & 0
\\
& &
\nonumber \\
\hat{\nabla}_{\mu}\left( e^{-2\phi} G_{mn} F^{(1)n\mu\alpha}\right)
& = & 0\, ,
\\
& &
\nonumber \\
\hat{\nabla}_{\mu}\left( e^{-2\phi}G^{mn}{\cal
F}_{n}{}^{\mu\alpha}\right)
& = & 0\, ,
\\
& &
\nonumber \\
\hat{\nabla}_{\mu}\left( e^{-4\phi}H^{\mu\alpha\beta}\right)
& = & 0\, ,
\label{eq:n4d4eqmolast}
\end{eqnarray}

\noindent where $T^{\phi}_{\alpha\beta}$ is the energy-momentum tensor
of $\phi$ (just as in Eq.~(\ref{eq:tscalar})) and $T^{B}_{\alpha\beta}$
is the energy-momentum tensor of the axion two-form $B_{\alpha\beta}$

\begin{equation}
T^{B}_{\alpha\beta}= H_{\alpha}{}^{\mu\nu}H_{\beta\mu\nu}
-{\textstyle\frac{1}{6}}\tilde{g}_{\alpha\beta}H^{2}\, .
\end{equation}

We also have to satisfy the following Bianchi identities

\begin{equation}
\begin{array}{rclrcl}
\partial F^{(1)m}
&
=
&
0\, ,
&
\partial H
&
=
&
{\textstyle\frac{1}{2}}F^{(1)m}F^{(2)}{}_{m}\, ,
\\
& & & & & \\
\partial F^{(2)}{}_{m}
&
=
&
0\, .
& & & \\
\end{array}
\label{eq:bianchis}
\end{equation}

Any truncation leading to the $a$-model must necessarily have no axion
field and no axion field-strength. A reasonable choice is, then

\begin{equation}
\begin{array}{rclrcl}
B_{\alpha\beta} & = & 0\, , & H_{\alpha\beta\gamma} & = & 0\, , \\
& & & & & \\
G_{mn} & = & -e^{2\rho_{m}}\delta_{mn}\, , & B_{mn} & = & 0\, .
\end{array}
\label{eq:truncation}
\end{equation}

Substituting into the above equations of motion and Bianchi identities
we get the following equations of motion

\begin{eqnarray}
\tilde{G}_{\alpha\beta} + 2T^{\phi}_{\alpha\beta} +
\sum_{m}T_{\alpha\beta}^{\rho_{m}}
& &
\nonumber \\
& &
\nonumber \\
-{\textstyle\frac{1}{2}}
\sum_{m}e^{-2(\phi-\rho_{m})}T_{\alpha\beta}^{(1)m}
-{\textstyle\frac{1}{2}}\sum_{m}
e^{-2(\phi+\rho_{m})}T_{m\alpha\beta}^{(2)}
& = & 0\, ,
\\
& &
\nonumber \\
\tilde{\nabla}^{2}\phi
-{\textstyle\frac{1}{8}}
\sum_{m}e^{-2(\phi-\rho_{m})} \left( F^{(1)m} \right)^{2}
-{\textstyle\frac{1}{8}}
\sum_{m} e^{-2(\phi-\rho_{m})} \left(F^{(2)}{}_{m} \right)^{2}
& = & 0\, ,
\\
& &
\nonumber \\
\tilde{\nabla}^{2}\rho_{m}
+{\textstyle\frac{1}{4}} e^{-2(\phi-\rho_{m})}\left(F^{(1)m} \right)^{2}
-{\textstyle\frac{1}{4}} e^{-2(\phi+\rho_{m})}\left(F^{(2)}{}_{m}
\right)^{2}
& = & 0\, ,
\\
& &
\nonumber \\
\tilde{\nabla}_{\mu}\left( e^{-2(\phi-\rho_{m})}F^{(1)m\mu\alpha}
\right) & = & 0\, ,
\\
& &
\nonumber \\
\tilde{\nabla}_{\mu}\left( e^{-2(\phi+\rho_{m})}
F^{(2)}{}_{m}{}^{\mu\alpha} \right)
& = & 0\, ,
\end{eqnarray}

\noindent and the following constraints

\begin{eqnarray}
F^{(1)r}{}_{\mu\nu}F^{(1)s\mu\nu}
-e^{-2(\rho_{r}+\rho_{s})}F^{(2)}{}_{r\mu\nu} F^{(2)}{}_{s\mu\nu}
& = & 0\, , \hspace{.5cm} \forall r \neq s\, ,
\\
& &
\nonumber \\
F^{(1)r}{}_{\mu\nu}F^{(2)}{}_{s}{}^{\mu\nu}
-e^{-2(\rho_{r}-\rho_{s})}F^{(1)s}{}_{\mu\nu} F^{(2)}{}_{s}{}^{\mu\nu}
& = & 0\, , \hspace{.5cm} \forall r \neq s\, ,
\\
& &
\nonumber \\
\sum_{m}F^{(1)m}{}_{\mu\nu}{}^{\star}F^{(2)}{}_{m\mu\nu}
& = & 0\, .
\end{eqnarray}

The origins of the first two constraints are the equations of motion of
the vanishing fields.  The last constraint is a consistency condition
between $B_{\alpha\beta}=0$ and $H_{\alpha\beta\gamma}=0$ due to the
Bianchi identity of $B_{\alpha\beta}$ (\ref{eq:bianchis}) and it simply
means that the Chern-Simons term in the definition of $H$ locally
vanishes.

The $a$-model has only one scalar and one vector field.  It is
necessary, then, to get to it, to make an ansatz of this kind:

\begin{equation}
\begin{array}{rclrcl}
\vec{F}^{(1)}
&
=
&
\vec{n} F + \vec{p}\ {}^{\star}F\, ,
&
\rho_{m}
&
=
&
c_{m}\ \varphi\, ,
\\
& & & & & \\
\vec{F}^{(2)} & = & \vec{m} F + \vec{q}\ {}^{\star}F\, , &
\phi & = & b \varphi\, ,
\end{array}
\end{equation}

\noindent where $b$ and the $c_{m}$'s are constants and the vectors
$\vec{n},\vec{p},\vec{m},\vec{q}$ can be functions of $\varphi$ and we
have arranged the vector field-strengths in column vectors. $F$ is a
purely electric or magnetic vector field-strength (for definiteness we
take it to be electric).

It is clear that, after the truncation Eqs.~(\ref{eq:truncation}), no
other ansatz will take us to the $a$-model or, at least, to the static
electric black-hole solutions of the $a$-model.

Substituting our ansatz into the equations of motion and constraints it
is possible to prove (after a considerable amount of work) that only the
cases $a=\sqrt{3},1,1/\sqrt{3},0$ can be obtained.  Only in these cases
can the $a$-model be embedded in the truncation of the $N=4 (+6V),D=4$
supergravity theory that we have proposed.  One also finds that, up to
{\it heterotic} duality rotations, there is a very small number of ways
to do this embedding in each case (see Table~\ref{tab-embeddings} for a
complete collection of these embeddings).  Before we explain our results
in each case let us say that $T$~duality acts in our truncated model by
rotations separately in the space of the $A^{(1)m}$ vectors and in the
space of $A^{(2)}{}_{m}$ and by the transformation (Buscher's
$T$~duality transformation \cite{kn:B})

\begin{equation}
\begin{array}{rclrcl}
A^{(1)m\prime}
&
=
&
-A^{(2)}{}_{m}\, ,
&
\rho_{m}^{\prime}
&
=
&
-\rho_{m}\, ,
\\
& & & & & \\
A^{(2)}{}_{m}^{\prime}
&
=
&
-A^{(1)m}\, .
& & & \\
\end{array}
\end{equation}

There are also electric-magnetic dualities, which are essentially those
of the $a$-model Eqs.~(\ref{eq:amemdual}).

All these {\it heterotic} dualities (which do not assume the existence
of isometries in the four-dimensional solutions, but make use of the
fact that, as ten-dimensional solutions they do have a six-dimensional
Abelian isometry group) are just non-compact symmetries of the
supergravity theories \cite{kn:MS} and are, as explained in the
Introduction, consistent with supersymmetry in the sense that a
configuration and its duals have the same number of four-dimensional
unbroken supersymmetries\footnote{This is not necessarily true for the
ten-dimensional supersymmetries \cite{kn:BBKOBSHAAGB}.  If the
ten-dimensional Killing spinor depends on a compact dimension, there
will not be a corresponding four-dimensional Killing spinor depending
only on the four non-compact space-time dimensions.  This can happen
even though all fields are assumed to be independent of the compact
dimensions.  Thus, one could find that the same solution, as a
ten-dimensional solution has more solutions than as a four-dimensional
solution.  However, we will not find this kind of ten-dimensional
Killing spinors here.} \cite{kn:O1,kn:O2,kn:BBKOBSHAAGB,kn:SS,kn:IPT}.
This fact allows us to study just one configuration and not its {\it
heterotic} duals.

As we are going to see (see also \cite{kn:R}) the number of inequivalent
extreme black-hole solutions is very small and one always can choose a
representative in the equivalence class which has a maximum of two
vector and two scalar fields different from zero.

%%%%%%%%%%%%%%%%%%%%%%%%%%%%%%%%%%%%%%%%%%%%%%%%%%%%%%%%%%%%%%%%%%%%%%

\subsection{$a=\protect\sqrt{3}$ embeddings}
\label{ssec-embasqrt3}

It is easy to see that setting

\begin{equation}
F^{(1)}=\sqrt{2}F\, ,\
\hspace{.5cm}
\phi = {\textstyle\frac{1}{\sqrt{3}}} \varphi\, ,
\hspace{.5cm}
\rho_{1}= -{\textstyle\frac{2}{\sqrt{3}}}\varphi\, ,
\end{equation}

\noindent the equations of motion of the $N=4(+6V),D=4$ theory
Eqs.~(\ref{eq:n4d4eqmofirst})-(\ref{eq:n4d4eqmolast}) reduce to those of
the $a$-model with $a=\sqrt{3}$ \cite{kn:HM}.  Taking then the
multi-black-hole solution in Eqs.~(\ref{eq:mbhasolutions}) for this
value of $a$ we get the following corresponding solution of
$N=4(+6V),D=4$ in the string frame entirely expressed in terms of the
harmonic function $V$:

\begin{equation}
\left\{
\begin{array}{rcl}
ds^{2} & = & V^{-1}dt^{2} -d\vec{x}^{2}\, , \\
& & \\
e^{2\phi} & = & V^{-\frac{1}{2}}\, ,\\
& & \\
G_{11} & = & -V\, , \\
& & \\
A^{(1)1}{}_{t} & = & n\ V^{-1}\, .
\end{array}
\right.
\end{equation}

Using Eqs.~(\ref{eq:uplifting}) of Appendix~\ref{sec-IIAsusy} we can
readily express it in ten-dimensional form:

\begin{equation}
\left\{
\begin{array}{rcl}
d\hat{s}^{2} & = & V^{-1}dt^{2} -d\vec{x}^{2}
-\left( V^{\frac{1}{2}} dx^{\underline{4}}
+n\ V^{-\frac{1}{2}}dt\right)^{2}
-dx^{\underline{I}} dx^{\underline{I}}\, , \\
& & \\
\hat{B} & = & \hat{\phi} = 0\, .\\
\end{array}
\right.
\end{equation}

This configuration is purely gravitational in ten dimensions and
corresponds to the Kaluza-Klein black hole, first identified as a
solution of heterotic string theory in the last of the references in
\cite{kn:HM}.  Its $T$~dual is known as the (electrically charged)
$H$-monopole \cite{kn:HM}.

\begin{equation}
\left\{
\begin{array}{rcl}
d\hat{s}^{2} & = & V^{-1}dt^{2} -d\vec{x}^{2}
-V^{-1}(dx^{\underline{4}})^{2}
-dx^{\underline{I}} dx^{\underline{I}}\, , \\
& & \\
\hat{B} & = & -n\ \left( V^{-\frac{1}{2}}dt \right) \wedge
\left( V^{-\frac{1}{2}}dx^{\underline{4}} \right)\, ,\\
& & \\
e^{2\hat{\phi}} & = & V^{-1}\, ,
\end{array}
\right.
\end{equation}

\noindent and has both dilaton and axion non-vanishing.

%%%%%%%%%%%%%%%%%%%%%%%%%%%%%%%%%%%%%%%%%%%%%%%%%%%%%%%%%%%%%%%%%%%%%%

\subsection{$a=1$ embeddings}
\label{ssec-emba1}

Setting

\begin{equation}
\phi=\varphi\, ,
\hspace{.5cm}
F^{(1)1}=F\, ,
\hspace{.5cm}
F^{(2)}{}_{1}=\mp F\, ,
\label{eq:a1embedding1}
\end{equation}

\noindent we get the $a=1$ model. The corresponding $N=4,D=4$ solution
in the string frame is

\begin{equation}
\left\{
\begin{array}{rcl}
ds^{2} & = & V^{-2}dt^{2} -d\vec{x}^{2}\, , \\
& & \\
e^{2\phi} & = & V^{-1}\, ,\\
& & \\
A^{(1)1}{}_{t} = \mp A^{(2)}{}_{1t} & = & n\ V^{-1}\, ,\\
\end{array}
\right.
\label{eq:a1embedding1b}
\end{equation}

\noindent and the corresponding $N=1(2),D=10$ solution is

\begin{equation}
\left\{
\begin{array}{rcl}
d\hat{s}^{2} & = & V^{-2}dt^{2} -d\vec{x}^{2}
-\left( dx^{\underline{4}} +n\ V^{-1}dt\right)^{2}\\
& & \\
& &
-dx^{\underline{I}} dx^{\underline{I}}\, , \\
& & \\
\hat{B} & = & \mp n\ V^{-1}dt\wedge \left(dx^{\underline{4}}
+n\ V^{-1}dt\right)\, , \\
& & \\
e^{2\hat{\phi}} & = & V^{-1}\, .\\
\end{array}
\right.
\label{eq:a1embedding1c}
\end{equation}

If we choose the minus sign, as explained in Appendix~\ref{sec-IIAsusy},
the matter vector field combination $F^{(1)1} +{\cal F}_{1} =F^{(1)1}
+F^{(2)}{}_{1}$ vanishes, and only the supergravity vector field
combination $F^{(1)1} -{\cal F}_{1} =F^{(1)1} -F^{(2)}{}_{1}$ remains.
The $a=1$ model can thus be embedded in the pure $N=4,D=4$ supergravity
theory.  This was the embedding proposed in Ref.~\cite{kn:KLOPP}, and,
as we will see in the next section, it is the embedding which admits
Killing spinors and unbroken $N=4,D=4$ supersymmetry (one half of it).

If we choose the plus sign, only the matter vector field combination
remains, and the resulting embedding does not have any $N=4(+6V),D=4$
unbroken supersymmetry \cite{kn:GHS}.

This result seems paradoxical, since, after all, the four-dimensional
solutions are identical, and Bogomol'nyi-type bounds must be saturated
in both cases.

All this was done in the framework of the $N=4(+6V), D=4$ theory which
results from dimensional reduction of the {\it positive chirality}
$N=1,D=10$ supergravity theory, which is the standard choice.  In $N=8$
supergravity we are forced to consider both chiralities and the apparent
paradox will be explained (see next section).

On top of the two embeddings (\ref{eq:a1embedding1}) there is another
embedding of the $a=1$ multi-black-hole solution \cite{kn:CYKK}:

\begin{equation}
F^{(1)1}=F\, ,
\hspace{.5cm}
F^{(1)2}=\pm e^{-2\varphi}{}^{\star}F\, ,
\hspace{.5cm}
\rho_{1}=-\varphi\, ,
\hspace{.5cm}
\rho_{1}=+\varphi\, .
\label{eq:a1embedding2}
\end{equation}

The corresponding $N=4(+6V),D=4$ solution in the string frame is

\begin{equation}
\left\{
\begin{array}{rcl}
ds^{2} & = & V^{-1}dt^{2} -Vd\vec{x}^{2}\, , \\
& & \\
G_{11} & = & -V\, , \\
& & \\
G_{22} & = & -V^{-1}\, , \\
& & \\
A^{(1)1}{}_{t} & = & n\ V^{-1}\, ,\\
& & \\
A^{(1)2}{}_{\underline{i}} & = & m\ V_{\underline{i}}\, ,
\end{array}
\right.
\label{eq:a1embedding2b}
\end{equation}

\noindent and the ten-dimensional solution is

\begin{equation}
\left\{
\begin{array}{rcl}
d\hat{s}^{2} & = & V^{-1}dt^{2} -Vd\vec{x}^{2}
-\left( V^{\frac{1}{2}} dx^{\underline{4}}
+n\ V^{-\frac{1}{2}}dt\right)^{2}\\
& & \\
& &
-\left( V^{-\frac{1}{2}} dx^{\underline{5}}
+m\ V^{-\frac{1}{2}} V_{\underline{i}} dx^{\underline{i}}\right)^{2}
-dx^{\underline{I}} dx^{\underline{I}}\, , \\
& & \\
\hat{B} & = & \hat{\phi} = 0\, .\\
\end{array}
\right.
\label{eq:a1embedding2c}
\end{equation}

%%%%%%%%%%%%%%%%%%%%%%%%%%%%%%%%%%%%%%%%%%%%%%%%%%%%%%%%%%%%%%%%%%%%%%

\subsection{$a=1/\protect\sqrt{3}$ embeddings}
\label{ssec-emba1sqrt3}

Setting

\begin{equation}
\begin{array}{rclrcl}
F^{(1)1} & = & {\textstyle\sqrt{\frac{2}{3}}}F\, , &
F^{(1)2} & = & \mp F^{(2)}{}_{2}={\textstyle\sqrt{\frac{2}{3}}}\
e^{-\frac{2}{\sqrt{3}}\varphi} {}^{\star}F\, , \\
& & & & & \\
\phi & = & -\frac{1}{\sqrt{3}}\varphi\, ,&
\rho_{1} & = & -\frac{2}{\sqrt{3}}\varphi\, , \\
\end{array}
\label{eq:a1sqrt3embedding}
\end{equation}

\noindent we get the $a=1/\sqrt{3}$ model and the corresponding solution
of the $N=4(+6V),D=4$ theory\footnote{Observe that the Hodge dual
${}^{\star}F$ in the above formulae has to be calculated in the Einstein
frame metric.}

\begin{equation}
\left\{
\begin{array}{rcl}
ds^{2} & = & V^{-1}dt^{2} -V^{2}d\vec{x}^{2}\, , \\
& & \\
e^{2\phi} & = & V^{\frac{1}{2}}\, ,\\
& & \\
G_{11} & = & -V\, , \\
& & \\
A^{(1)1}{}_{t} & = & n\ V^{-1}\, ,\\
& & \\
A^{(1)2}{}_{\underline{i}} & = & \mp A^{(2)}{}_{2\underline{i}} = n
V_{\underline{i}}\, ,
\end{array}
\right.
\label{eq:a1sqrt3embeddingb}
\end{equation}

\noindent where $V_{\underline{i}}$ satisfies the equation

\begin{equation}
\partial_{[\underline{i}}V_{\underline{j}]} = {\textstyle\frac{1}{2}}
\epsilon_{ijk}\partial_{\underline{k}}V\, .
\end{equation}

The corresponding solution of the $N=1(2),D=10$ theory is

\begin{equation}
\left\{
\begin{array}{rcl}
d\hat{s}^{2} & = & V^{-1}dt^{2}-V^{2}d\vec{x}^{2}
-\left( V^{\frac{1}{2} }dx^{\underline{4}} +n\
V^{-\frac{1}{2}}dt \right)^{2} \\
& & \\
& &
-\left(dx^{\underline{5}}
+n\ V_{\underline{i}}dx^{\underline{i}} \right)^{2}
-dx^{\underline{I}} dx^{\underline{I}}\, , \\
& & \\
\hat{B} & = & \mp n\ V_{\underline{i}}dx^{\underline{i}} \wedge
\left(dx^{\underline{5}} +n\ V_{\underline{i}}dx^{\underline{i}}
\right)\, ,\\
& & \\
e^{2\hat{\phi}} & = & V\, .
\end{array}
\right.
\label{eq:a1sqrt3embeddingc}
\end{equation}

In the framework explained at the beginning of this section, no other
embedding of the $a=1/\sqrt{3}$ multi-black-hole solution is possible.

%%%%%%%%%%%%%%%%%%%%%%%%%%%%%%%%%%%%%%%%%%%%%%%%%%%%%%%%%%%%%%%%%%%%%%

\subsection{$a=0$ embeddings}
\label{ssec-emba0}

Setting

\begin{equation}
\begin{array}{rclrcl}
F^{(1)1} & = & {\textstyle\frac{1}{2}}F\, , &
F^{(1)2} & = & {\textstyle\frac{1}{2}}{}^{\star}F\, , \\
& & & & & \\
F^{(2)}{}_{1} & = & \mp{\textstyle\frac{1}{2}}F\, , &
F^{(2)}{}_{2} & = & \mp{\textstyle\frac{1}{2}}{}^{\star}F\, , \\
\end{array}
\label{eq:a0embedding}
\end{equation}

\noindent one gets the $a=0$ model (Einstein-Maxwell theory).  The
solution of the $N=4(+6V),D=4$ theory is

\begin{equation}
\left\{
\begin{array}{rcl}
ds^{2} & = & V^{-2}dt^{2} -V^{2}d\vec{x}^{2}\, , \\
& & \\
A^{(1)1}{}_{t} = \mp A^{(2)}{}_{1t} & = & n\ V^{-1}\, ,\\
& & \\
A^{(1)2}{}_{\underline{i}} = \mp A^{(2)}{}_{2\underline{i}} & = &
m\ V_{\underline{i}}\, ,\\
\end{array}
\right.
\label{eq:a0embeddingb}
\end{equation}

\noindent and uplifted to ten dimensions is

\begin{equation}
\left\{
\begin{array}{rcl}
d\hat{s}^{2} & = & V^{-2}dt^{2}-V^{2}d\vec{x}^{2}
-\left( dx^{\underline{4}} +n\ V^{-1}dt \right)^{2} \\
& & \\
& &
-\left(dx^{\underline{5}}
+m\ V_{\underline{i}}dx^{\underline{i}} \right)^{2}
-dx^{\underline{I}} dx^{\underline{I}}\, , \\
& & \\
\hat{B} & = & \mp n\ V^{-1}dt\wedge \left(dx^{\underline{4}} +n\
V^{-1}dt \right)\\
& & \\
& &
\mp m\ V^{-1} V_{\underline{i}}\left(V dx^{\underline{i}}\right)
\wedge \left(dx^{\underline{5}} +m\ V_{\underline{i}}dx^{\underline{i}}
\right)\, ,\\
\end{array}
\right.
\label{eq:a0embeddingc}
\end{equation}

With the minus sign, this is the standard embedding of the extreme
Reissner-Nordstr\"om solution into $N=4(+6),D=4$ supergravity
\cite{kn:KLOPP}.  We can repeat here the discussion we made in the $a=1$
case with respect to the relative sign of $F^{(1)m}$ and $F^{(2)}{}_{m}$
and the matter or supergravity nature of the vector fields.

No other purely electric or magnetic embeddings exist (up to dualities),
but there exist some other (dyonic) embeddings \cite{kn:KO} that we are
going to discuss now\footnote{Observe that for the other values of $a$
no dyonic embedding of any kind exists.  This reflects two facts: (i)
there are no dyonic solution of the $a$-model for $a\neq 0,1$, and (ii)
the dyonic solution of the $a=1$ model is clearly not a solution of the
$N=4,D=4$ theory, as explained in Ref.~\cite{kn:KLOPP}}.

%%%%%%%%%%%%%%%%%%%%%%%%%%%%%%%%%%%%%%%%%%%%%%%%%%%%%%%%%%%%%%%%%%%%%%

\subsubsection{Dyonic embeddings}

The simplest of these embeddings is the following:

\begin{equation}
F^{(1)1} = F\pm {}^{\star}F\, .
\end{equation}

The essential property that makes this embedding a solution of the
$N=4(+6V),D=4$ theory is that, given that $F$ is purely electric or
magnetic $F{}^{\star}F=0$ and, then $\left( F^{(1)1} \right)^{2}=0$.
The solution is, thus,

\begin{equation}
\left\{
\begin{array}{rcl}
ds^{2} & = & V^{-2}dt^{2} -V^{2}d\vec{x}^{2}\, , \\
& & \\
A^{(1)1}& = & \sqrt{2} \left( n\ V^{-1}dt +m\
V_{\underline{i}}dx^{\underline{i}} \right)\, ,\\
\end{array}
\right.
\end{equation}

\noindent and the corresponding ten-dimensional solution is

\begin{equation}
\left\{
\begin{array}{rcl}
d\hat{s}^{2} & = & V^{-2}dt^{2} -V^{2}d\vec{x}^{2}
-\left[ dx^{\underline{4}}
+\sqrt{2}\left(n\ V^{-1}dt+ m\ V_{\underline{i}}
dx^{\underline{i}}\right)\right]^{2}\\
& & \\
& &
-dx^{\underline{I}}dx^{\underline{I}}\, , \\
& & \\
\hat{B} & = & \hat{\phi} = 0\, .\\
\end{array}
\right.
\end{equation}

This embedding can be generalized to the form

\begin{equation}
\vec{F}^{(1)}=\vec{n}  \left( F\pm {}^{\star}F \right),
\qquad\qquad  \vec{F}^{(2)}=\vec{m}  \left( F\mp {}^{\star}F \right)\, ,
\end{equation}

\noindent provided $n^rn^r+m^rm^r=1$ and $n^rm^s=n^sm^r$. Some of these
embeddings are just $T$~or $S$~duals of the simplest one, but we will
not pursue this problem here.

%%%%%%%%%%%%%%%%%%%%%%%%%%%%%%%%%%%%%%%%%%%%%%%%%%%%%%%%%%%%%%%%%%%%%%

\section{Unbroken $N=4(8)$ supersymmetries of the $a$-model solutions}
\label{sec-supersymmetry}

In the previous section we have found solutions of $N=1(2),D=10$
supergravity which in four dimensions correspond to the multi-black-hole
solutions of the $a$-model.  In this section we are going to find the
unbroken supersymmetries of the ten-dimensional solutions, which is a
way of finding the unbroken supersymmetries of the four dimensional
solutions in $N=4(8)$ supergravity.

As explained in Appendix~\ref{sec-IIAsusy}, when the R-R
fields of the type~IIA theory vanish, the supersymmetry rules reduce to
Eqs.~(\ref{eq:norrrules}) that we rewrite here for convenience

\begin{eqnarray}
\delta_{\epsilon} \hat{\psi}^{(\pm)}_{\hat{a}} & = &
\hat{\nabla}_{\hat{a}}^{(\pm)}\ \hat{\epsilon}^{(\pm)}\, ,
\nonumber \\
& &
\nonumber \\
\delta_{\epsilon} \hat{\lambda}^{(\pm)} & = &
\left(\not\!\partial\hat{\phi}
\pm {\textstyle\frac{1}{4}}\not\!\!\hat{H}\right)
\hat{\epsilon}^{(\pm)}\, ,
\nonumber
\end{eqnarray}

\noindent where $\hat{\nabla}_{\hat{a}}^{(\pm)}$ are the covariant
derivatives associated to the two torsionful spin connections

\begin{displaymath}
\hat{\Omega}^{(\pm)}_{\hat{a} \hat{b} \hat{c}}
=\hat{\omega}_{\hat{a} \hat{b} \hat{c}}
\mp{\textstyle\frac{3}{2}} \hat{H}_{\hat{a} \hat{b} \hat{c}}\, .
\end{displaymath}

\noindent Taking just the positive chirality (upper signs) in the above
equations one gets the supersymmetry rules of the conventional
$N=1,D=10$ supergravity theory $N=1^{(+)},D=10$.  The other sign choice
gives the supersymmetry rules of another $N=1,D=10$ theory
($N=1^{(-)},D=10$) that can be constructed.

In fact, to find unbroken supersymmetries, in many cases we can use the
symmetry existing between the two chirality sectors of the NS-NS sector
of type~IIA supergravity: ``$C$~duality''.  A $C$~duality transformation
changes the chirality of the spinors and the sign of the axion and
leaves the NS-NS sector of the type~IIA theory invariant.  However, from
the point of view of the $N=1$ theories, $C$~duality is not a symmetry.
Each theory has a definite chirality that cannot be changed.  Instead,
$C$~duality takes us from the $N=1^{(+)},D=10$ theory to the
$N=1^{(-)},D=10$ theory and is another (very simple) example of
string/string duality.

Then, when we have two embeddings which differ by just the sign of the
axion, we can use $C$~duality arguments to translate the results of one
chirality sector to the other.

The necessary ingredients to find the unbroken supersymmetries are just
the dilaton, the axion field strength and the spin connection
coefficients, which are calculated for the kind of metric we are dealing
with in Appendix~\ref{sec-spincon}.

%%%%%%%%%%%%%%%%%%%%%%%%%%%%%%%%%%%%%%%%%%%%%%%%%%%%%%%%%%%%%%%%%%%%%%

\subsection{Unbroken supersymmetries of the $a=\protect\sqrt{3}$
embeddings}

The supersymmetry equations for $a=\sqrt{3}$ black holes are the most
straightforward.  Consider the field configuration of the electrically
charged Kaluza-Klein black hole described in
Section~\ref{ssec-embasqrt3}.  Then from $\hat{\phi} =\not\!\!\hat{H}
=0$, it follows that the supersymmetry rules are identical for both
positive and negative chirality ten-dimensional spinors and that the
dilatino equation is trivially satisfied.  Following
Appendix~\ref{sec-IIAsusy}, it is straightforward to show that the
Killing spinors of this embedding are those which satisfy

\begin{eqnarray}
\hat{\epsilon}^{(\pm)} & = &
V^{-\frac{1}{4}}\ \hat{\epsilon}^{(\pm)}_{(0)}\, ,\\
& & \nonumber \\
\hat{\gamma}_{0} \hat{\gamma}_{4}\ \hat{\epsilon}^{(\pm)}_{(0)} & = &
-n\ \hat{\epsilon}^{(\pm)}_{(0)}\, ,
\end{eqnarray}

\noindent where $\hat{\epsilon}^{(\pm)}_{(0)}$ is a constant spinor.
This chirality condition on the subspinor in the $(04)$ sector implies
that precisely half of the supersymmetries are broken, for either
positive or negative ten-dimensional spinor.  So half of the
supersymmetries are preserved for each of the opposite chirality $N=1$
theories in $D=10$, and, as a result, half of each of the corresponding
$N=4$ supersymmetries in $D=4$, and, therefore, a total of a half of the
$N=8,D=4$ supersymmetries.

Had we performed the same calculation for the $T$~dual of this solution,
namely the electrically charged $H$-monopole, which has a non-vanishing
axion field strength and is not $C$~duality invariant, we would have
gotten a similar result:

\begin{eqnarray}
\hat{\epsilon}^{(\pm)} & = &
V^{-\frac{1}{4}}\ \hat{\epsilon}^{(\pm)}_{(0)}\, ,\\
& & \nonumber \\
\hat{\gamma}_{0} \hat{\gamma}_{4}\ \hat{\epsilon}^{(\pm)}_{(0)} & = &
\mp n\ \hat{\epsilon}^{(\pm)}_{(0)}\, .
\end{eqnarray}

As can be seen above, the Killing spinors are invariant under
$T$~duality in the positive chirality sector, but are only covariant in
the negative chirality sector.  This is due to the different ways in
which the two torsionful spin connections $\hat{\Omega}^{(\pm)}$
transform \cite{kn:BEK}.  The number of $N=4(8),D=4$ unbroken
supersymmetries does not change, though.  It would also be the same had
we taken the magnetically charged $S$~dual versions of both the
Kaluza-Klein and $H$-monopole, although in these latter two cases, the
chirality condition is imposed on the $(1234)$ sector of the spinor.
Henceforth we will not consider explicitly the $S$~or $T$~dual versions
of these solutions.

%%%%%%%%%%%%%%%%%%%%%%%%%%%%%%%%%%%%%%%%%%%%%%%%%%%%%%%%%%%%%%%%%%%%%%

\subsection{Unbroken supersymmetries of the $a=1$ embeddings}

The situation for the $a=1$ embeddings is a bit more subtle.  Consider
the first embedding in Section~\ref{ssec-emba1},
Eqs.~(\ref{eq:a1embedding1}-\ref{eq:a1embedding1c}) and the positive
(conventional) chirality $N=1^{(+)},D=10$ theory.  When the minus sign
is chosen, both the gravitino and dilatino equations reduce to the same
conditions, namely

\begin{eqnarray}
\hat{\epsilon}^{(+)} & = &
V^{-\frac{1}{2}}\ \hat{\epsilon}^{(+)}_{(0)}\, ,\\
& & \nonumber \\
\hat{\gamma}_{0} \hat{\gamma}_{4}\ \hat{\epsilon}^{(+)}_{(0)} & = &
-n\ \hat{\epsilon}^{(+)}_{(0)}\, .
\end{eqnarray}

When the plus sign is chosen, there are no Killing spinors in the
$N=1^{(+)},D=10$ theory ({\it i.e.}~$\hat{\epsilon}^{(+)}=0$ is the
only consistent solution).

As explained above, these two choices of sign correspond respectively to
declaring that the four dimensional vector is a supergravity vector (and
that the matter vector vanishes) in the $N=4,D=4$ theory and declaring
exactly the opposite.  We have just reproduced the results of
Refs.~\cite{kn:KLOPP,kn:GHS} respectively, although in a different
setting.

We can now use $C$~duality to find the unbroken supersymmetries in the
$N=1^{(-)},D=10$ theory.  For the minus sign choice there is now no
Killing spinor, and for the plus sign one gets the same condition as for
the minus sign in the $N=1^{(+)},D=10$ theory, namely

\begin{eqnarray}
\hat{\epsilon}^{(-)} & = &
V^{-\frac{1}{2}}\ \hat{\epsilon}^{(-)}_{(0)}\, ,\\
& & \nonumber \\
\hat{\gamma}_{0} \hat{\gamma}_{4}\ \hat{\epsilon}^{(-)}_{(0)} & = &
-n\ \hat{\epsilon}^{(-)}_{(0)}\, .
\end{eqnarray}

Then, both choices of sign (both embeddings) are supersymmetric in one
sector, and in that sector a half of the $N=4,D=4$ supersymmetries are
unbroken, just as in the $a=\sqrt{3}$ case.  Since we are forced to
consider both sectors, the total number of $N=8$ unbroken
supersymmetries is the same for both choices: $1/4$.  This result, which
resolves the paradox, could also be explained by the fact that there are
no matter vector fields in the $N=8$ theory: {\it all} vectors are
supergravity vector fields.

Finally, consider the embedding of
Eqs.~(\ref{eq:a1embedding2}-\ref{eq:a1embedding2c}).  This case, as the
$a=\sqrt{3}$ (Kaluza-Klein) case, is $C$~duality symmetric, and the
supersymmetry transformations are identical for both chiralities.  In
addition to the conditions

\begin{eqnarray}
\hat{\epsilon}^{(\pm)} & = & V^{-\frac{1}{4}}\
\hat{\epsilon}^{(\pm)}_{(0)}\, ,\\
& & \nonumber \\
\hat{\gamma}_{0} \hat{\gamma}_{4}\ \hat{\epsilon}^{(\pm)} & = &
-n\ \hat{\epsilon}^{(\pm)} \, ,
\end{eqnarray}

\noindent we get the condition

\begin{equation}
\hat{\Gamma}_{5}\ \hat{\epsilon}^{\pm}= -m\ \hat{\epsilon}^{\pm} \, ,
\end{equation}

\noindent coming from the magnetic sector, where $\hat{\Gamma}_{5}
=\hat{\gamma}_{1} \hat{\gamma}_{2} \hat{\gamma}_{3} \hat{\gamma}_{5}$.
As a result, the supersymmetries already halved by the first condition
are halved again in each sector.  This implies that for each $N=4$
theory, one quarter of the supersymmetries are preserved
\cite{kn:CYKK,kn:CYN8}.
Therefore, again, but in a different fashion, for the $N=8$ theory, $2$
supersymmetries (one quarter) are preserved.

%%%%%%%%%%%%%%%%%%%%%%%%%%%%%%%%%%%%%%%%%%%%%%%%%%%%%%%%%%%%%%%%%%%%%%

\subsection{Unbroken supersymmetries of the $a=1/\protect\sqrt{3}$
embeddings}

In some sense the $a=1/\sqrt{3}$ black hole is a combination of an
$a=1$ black hole with a dual-charged $a=\sqrt{3}$ black hole. Again
following Appendix~\ref{sec-spincon}, it is straightforward to show
that for the minus sign choice in the embedding of
Eqs.~(\ref{eq:a1sqrt3embedding}-\ref{eq:a1sqrt3embeddingc})

\begin{eqnarray}
\hat{\epsilon}^{(\pm)} & = & V^{-\frac{1}{4}}\
\hat{\epsilon}^{(\pm)}_{(0)}\, , \\
& & \nonumber \\
\hat{\Gamma}_{5}\ \hat{\epsilon}^{(+)}_{(0)} & = &
-n\ \hat{\epsilon}^{(+)}_{(0)} \, , \\
& & \nonumber \\
\hat{\gamma}_{0} \hat{\gamma}_{4}\ \hat{\epsilon}^{(+)}_{(0)} & = & -n\
\hat{\epsilon}^{(+)}_{(0)}\, .
\end{eqnarray}

\noindent and $\hat{\epsilon}^{(-)}=0$.  This implies that $1/4$ of the
positive chirality $N=4$ supersymmetries (i.e. the supersymmetries
arising from the reduction of the positive chirality ten-dimensional
spinor) are preserved while none of the negative chirality
supersymmetries are preserved.  As a result, only one of the $N=8$
supersymmetries is preserved.

For the plus sign choice in
Eqs.~(\ref{eq:a1sqrt3embedding}-\ref{eq:a1sqrt3embeddingc}), none of the
positive chirality supersymmetries are preserved, while one of the four
negative chirality supersymmetries is preserved, explicitly

\begin{eqnarray}
\hat{\epsilon}^{(-)} & = & V^{-\frac{1}{4}}\
\hat{\epsilon}^{(-)}_{(0)}\, , \\
& & \nonumber \\
\hat{\Gamma}_{5}\ \hat{\epsilon}^{(-)}_{(0)} & = &
+n\ \hat{\epsilon}^{(-)}_{(0)} \, , \\
& & \nonumber \\
\hat{\gamma}_{0} \hat{\gamma}_{4}\ \hat{\epsilon}^{(-)}_{(0)} & = & +n\
\hat{\epsilon}^{(-)}_{(0)}\, .
\end{eqnarray}

A simple way of seeing this pattern is as follows: the configuration
described in Eqs.~(\ref{eq:a1sqrt3embedding}-\ref{eq:a1sqrt3embeddingc})
represent a combination of a magnetic $a=1$ black hole and an electric
$a=\sqrt{3}$ black hole.\footnote{It is interesting that this composite
viewpoint is consistent with the bound state picture in \cite{kn:R} at
the level of the Killing spinor equations.} The $a=1$ part of the
configuration preserves half the supersymmetries for one chirality and
none for the other.  The $a=\sqrt{3}$ part then independently halves
again whatever remaining supersymmetries exist in each sector.  As a
result, we are left with only an eighth of the $N=8$ supersymmetries.

%%%%%%%%%%%%%%%%%%%%%%%%%%%%%%%%%%%%%%%%%%%%%%%%%%%%%%%%%%%%%%%%%%%%%%

\subsection{Unbroken supersymmetries of the $a=0$ embeddings}

The $a=0$ embeddings described in Section~\ref{ssec-emba0},
Eqs.~(\ref{eq:a0embedding}-\ref{eq:a0embeddingc}) have precisely the
same supersymmetry breaking pattern as the $a=1/\sqrt{3}$ solutions
described above, with the difference that an $a=0$ solution can most
usefully be thought of as a combination of two $a=1$ black holes, one
electric and one magnetic, each imposing an independent chirality
condition.

The dyonic embeddings we have shown, by contrast, break all of the
spacetime supersymmetries.  These embeddings do not, however, exhaust
all possible dyonic embeddings.  This can be seen by noting that there
exist dyonic ERN black holes which preserve some supersymmetry in
certain $N=2$ truncations\footnote{The purely electric and purely
magnetic ERN black holes are supersymmetric in $N=2,D=4$ supergravity.
This theory has an electric-magnetic duality symmetry that preserves
unbroken supersymmetries (see, for instance, the second lecture in
Ref.~\cite{kn:GWG}), and, therefore, the dyonic ERN black hole is
supersymmetric in that theory, which can be obtained by a consistent
truncation of $N=8,D=4$ supergravity.}.

%%%%%%%%%%%%%%%%%%%%%%%%%%%%%%%%%%%%%%%%%%%%%%%%%%%%%%%%%%%%%%%%%%%%%%

\section{Conclusion}
\label{sec-conclusion}

In this paper we have determined which four-dimensional extreme dilaton
black-hole solutions can be embedded in $N=4$ supergravity, for which
values of the parameter $a$ and in how many inequivalent ways this can
be done (that is, not related by duality symmetries).  We have also
studied the $N=4$ unbroken supersymmetries of these black holes as well
as their $N=8$ unbroken supersymmetries, making use of the fact that the
$N=4$ theory can be considered as a consistent truncation of the $N=8$.
Our results are summarized in Table~\ref{tab-embeddings}.

\begin{table}
\begin{center}
\begin{tabular}{||c||c|c|c|c|c|c|c||c||}
\hline\hline
& & & & & & & & \\
$a$ & $\phi$ & $\rho_{1}$ & $\rho_{2}$ &
$F^{(1)1}$ & $F^{(2)}{}_{1}$ & $F^{(1)2}$ & $F^{(2)}{}_{2}$ &
$(n_{+},n_{-})$ \\
\hline\hline
& & & & & & & & \\
$\sqrt{3}$ & $\frac{1}{\sqrt{3}}\varphi$ &
$-\frac{2}{\sqrt{3}}\varphi$ & $0$ &
$\sqrt{2}F$ & $0$ & $0$ & $0$ & $(\frac{1}{2},\frac{1}{2})$
\\
\hline\hline
& & & & & & & & \\
$1$ & $\varphi$ & $0$ & $0$ & $F$ & $-F$ & $0$ & $0$ & $(\frac{1}{2},0)$
\\
\cline{2-9}
& & & & & & & & \\
 & $\varphi$ & $0$ & $0$ & $F$ & $+F$ & $0$ & $0$ & $(0,\frac{1}{2})$
\\
\cline{2-9}
& & & & & & & & \\
 & $0$ & $-\varphi$ & $\varphi$ & $F$ & $0$ & $e^{-2\varphi}{}^{\star}F$
& $0$ & $(\frac{1}{4},\frac{1}{4})$
\\
\hline\hline
& & & & & & & & \\
$\frac{1}{\sqrt{3}}$ & $-\frac{1}{3}\varphi$ & $-\frac{2}{3}\varphi$ &
$0$ & $\sqrt{\frac{2}{3}}F$ & $0$ &
$\sqrt{\frac{2}{3}}e^{2\phi}{}^{\star}F$
& $-\sqrt{\frac{2}{3}}e^{2\phi}{}^{\star}F$ & $(\frac{1}{4},0)$
\\
\cline{2-9}
& & & & & & & & \\
 & $-\frac{1}{3}\varphi$ & $-\frac{2}{3}\varphi$ &
$0$ & $\sqrt{\frac{2}{3}}F$ & $0$ &
$\sqrt{\frac{2}{3}}e^{2\phi}{}^{\star}F$
& $+\sqrt{\frac{2}{3}}e^{2\phi}{}^{\star}F$ & $(0,\frac{1}{4})$
\\
\hline\hline
& & & & & & & & \\
$0$ & $0$ & $0$ & $0$ & $\frac{1}{\sqrt{2}}F$ &
$-\frac{1}{\sqrt{2}}F$ & $\frac{1}{\sqrt{2}}{}^{\star}F$ &
$-\frac{1}{\sqrt{2}}{}^{\star}F$ & $(\frac{1}{4},0)$
\\
\cline{2-9}
& & & & & & & & \\
 & $0$ & $0$ & $0$ & $\frac{1}{\sqrt{2}}F$ &
$+\frac{1}{\sqrt{2}}F$ & $\frac{1}{\sqrt{2}}{}^{\star}F$ &
$+\frac{1}{\sqrt{2}}{}^{\star}F$ & $(0,\frac{1}{4})$
\\
\cline{2-9}
& & & & & & & & \\
 & $0$ & $0$ & $0$ & $F\pm {}^{\star}F$ & $0$ & $0$ & $0$ & $(0,0)$
\\
\hline\hline
\end{tabular}
\end{center}

\caption[Table of embeddings and supersymmetries]
{In this table we give the different embeddings (up to $N=4$ (heterotic)
dualities) of the $a=\sqrt{3},1,1/\sqrt{3},0$ purely electric (or
magnetic) solutions Eqs.~(\ref{eq:mbhasolutions}) in $N=4(8)$
supergravity.  It is read in the following manner: if the $N=4$ fields of the
top row take the values given in the following rows, in terms of
$\varphi$ and $F$, where $F$ is either purely electric or purely
magnetic, then the $N=4$ equations of motion reduce to those of the
$a$-model (\ref{eq:aeqmo}) for the value of $a$ given in the first
column.  In the last column we list the unbroken supersymmetry in the
two $N=4$ sectors of positive and negative chirality as a fraction of
the total.}
\label{tab-embeddings}
\end{table}

We have found that only the $a=\sqrt{3}, 1, 1/\sqrt{3}, 0$ dilaton black
holes can be embedded in the $N=4$ theory and that this can be done in a
very limited number of inequivalent ways (not related by $T$~or
$S$~duality).  There is only one embedding of the $a=\sqrt{3}$ dilaton
black hole, three of the $a=1$ one and two of the $a=1/\sqrt{3}$.  The
$a=0$ can be embedded in just two different (purely electric or
magnetic) ways, but other (dyonic) embeddings are possible.  All the
inequivalent embeddings in the $N=4$ theory have different amounts of
unbroken supersymmetry.

The situation changes when we consider the embeddings in the $N=8$
theory: {\it all embeddings of the same dilaton black hole are
equivalent under $U$~duality and have the same number of unbroken
supersymmetries} with the exception of the dyonic embedding of the $a=0$
extreme black hole.  There are $U$~duality transformations that relate
embeddings which are inequivalent in the $N=4$ theory and do not change
the number of $N=8$ unbroken supersymmetries but do change the number of
$N=4$ unbroken supersymmetries, essentially by shifting the unbroken
supersymmetries from one chirality sector to the other.  One example is
the $C$~duality transformation that interchanges the two chirality
sectors and supergravity and matter fields of a given $N=4$ theory (all
vectors are supergravity vectors in the $N=8$ theory and this is why
$C$~duality {\it is} a symmetry of this theory).

Note that our analysis applies to the string-like solitons constructed
in \cite{kn:SS}, where some solutions were found to preserve some
supersymmetry provided one made a chirality choice that matched the
overall chirality of the $N=1, D=10$ theory from which the $N=4$ theory
was reduced.  On making the opposite chirality choice, however, it was
found that the solution broke all supersymmetries.  This also presented
an apparent paradox, since both embeddings represent essentially
identical solutions with analogous Bogomol'nyi bounds.  From the results
in this paper, however, it follows immediately that the ``wrong''
chirality choice embedding simply corresponds to a solution which
preserves supersymmetry in the opposite chirality $N=1, D=10$ theory.
In the $N=8$ theory, both chirality choices lead to embeddings which
preserve the same amount of supersymmetry.  This conclusion also applies
to analyticity versus anti-analyticity conditions in certain $N=1, D=4$
truncations \cite{kn:SS}.

Furthermore, our analysis in this paper can be generalized in a
straightforward manner to arbitrary supersymmetric $p$-branes, both
isotropic and anisotropic, in arbitrary $D$ spacetime dimensions,
following the oxidation/reduction procedures discussed in \cite{kn:KM2}
(see also \cite{kn:LPSS}).

It is tempting to conclude that all embeddings of any given
four-dimensional solution should be equivalent in the $N=8$ theory.
Previously it was thought that only special embeddings of a solution in
a supergravity theory had unbroken supersymmetry.  Our results seem to
indicate that if a solution saturates certain bounds and there is one
supersymmetric embedding, all possible embeddings will also be
supersymmetric, and none of them will be singled out.

This hypothesis could explain why we have found no embeddings with
$(\frac{1}{8},\frac{1}{8})$ of unbroken supersymmetries, that is, with
$\frac{1}{8}$ of the $N=8$ supersymmetries unbroken, half of them in
each chirality sector. $U$~duality transformations can only change the
number of unbroken supersymmetries by an integer number of $N=8$
supersymmetries.  Thus, if we start with the $(\frac{1}{4},0)$ embedding
of the $a=1/\sqrt{3}$ black hole, we can only get to the
$(0,\frac{1}{4})$ embedding, by using a $U$~duality transformation that
shifts one $N=8$ supersymmetry from the positive to the negative
chirality sector.  If our hypothesis is true, then, we cannot access
this embedding by $U$~duality, and it does not exist (certainly we have
not found it).

However, we cannot ignore the presence of a manifest exception to this
hypothesis: the dyonic embedding of the extreme Reissner-Nordstr\"{o}m
black hole.  An explanation of the existence of this solution in terms
of bound states is not apparent.  Instead, one could hope for a
larger framework in which this embedding is supersymmetric, just as
embeddings which are non-supersymmetric in the $N=4$ picture are
supersymmetric in the $N=8$ framework \cite{kn:KO}.

In calculating the $N=8$ unbroken supersymmetries we have used
the ten-dimensional type~IIA theory. Since we are considering
four-dimensional solutions, our results (the number of $N=8$ unbroken
supersymmetries) would be identical  had we worked with
the type~IIB theory. It is, though, of some interest, to know what the
ten-dimensional Killing spinors would look like in the type~IIB case,
since this theory is chiral and it has spinors of only one chirality.
Now one has two sectors of the same chirality.

The chirality of the type~IIB theory is conventionally positive, so one
gets the positive chirality $N=1^{(+)}$ theory upon truncation of the
bosonic RR fields and one of the spinor sectors (say the second).  But
there is also a negative chirality type~IIB theory (type~IIB${}^{(-)}$)
characterized by the different chirality of the spinors and by the fact
that the five-form $\hat{F}_{\hat{\mu}_{1} \ldots \hat{\mu}_{5}}$ (which
is the field-strength of the four-form field $\hat{D}_{\hat{\mu}_{1}
\ldots \hat{\mu}_{4}}$ in the notation of Refs.~\cite{kn:BHO,kn:BBO})
instead of being self-dual as in the type~IIB${}^{(+)}$ theory, is
anti-self-dual.  The same truncation of this theory would give us the
$N=1^{(-)}$ one.

The situation is summarized in Table~\ref{tab-sectors}.

It takes, then, little thought to arrive at the conclusion that, since
both type~II theories describe the same degrees of freedom but are
``arranged'' in different ways, there must be a supersymmetric basis for
the type~IIB${}^{(+)}$ theory such that, in absence of bosonic R-R
fields, the sector corresponding to the gravitino
$\hat{\psi}^{(+)1}_{\hat{a}}$ is the same as in the type~IIA${}^{1}$
theory and the sector corresponding to the gravitino
$\hat{\psi}^{(+)2}_{\hat{a}}$ is the same as the sector
$\hat{\psi}^{(-)1}_{\hat{a}}$ of the type~IIA${}^{1}$ theory.  From this
viewpoint, then, in the absence of R-R fields, the supersymmetry
transformation rules should be

\begin{equation}
\begin{array}{rclrcl}
\delta_{\epsilon} \hat{\psi}^{(+)1}_{\hat{a}} & = &
\hat{\nabla}_{\hat{a}}^{(+)}\ \hat{\epsilon}^{(+)1}\, , &
\delta_{\epsilon} \hat{\lambda}^{(+)1} & = &
\left(\not\!\partial\hat{\phi}
+ {\textstyle\frac{1}{4}}\not\!\!\hat{H}\right)
\hat{\epsilon}^{(+)1}\, , \\
& & & & & \\
\delta_{\epsilon} \hat{\psi}^{(+)2}_{\hat{a}} & = &
\hat{\nabla}_{\hat{a}}^{(-)}\ \hat{\epsilon}^{(+)2}\, , &
\delta_{\epsilon} \hat{\lambda}^{(+)2} & = &
\left(\not\!\partial\hat{\phi}
-{\textstyle\frac{1}{4}}\not\!\!\hat{H}\right)
\hat{\epsilon}^{(+)2}\, . \\
\end{array}
\end{equation}

\noindent In this case, all our results for the Killing spinors in the
type~IIA theory can be translated to the type~IIB${}^{(+)}$ by just
replacing $\hat{\epsilon}^{(+)}$ by $\hat{\epsilon}^{(+)1}$ and
$\hat{\epsilon}^{(-)}$ by $\hat{\epsilon}^{(+)2}$.  In the
type~IIB${}^{(+)}$ $C$~duality would relate the $1$ and $2$ sectors
$\hat{\psi}_{\hat{a}}^{(+)1}$ and $\hat{\psi}_{\hat{a}}^{(+)2}$ which
now happen to have the same chirality.  In the $N=1$ context,
$C$~duality would relate two different $N=1^{(+)}$ theories of the same
chirality, but with different supersymmetry rules, the difference being
the sign of $\hat{H}_{\hat{\alpha}\hat{\beta}\hat{\gamma}}$.

Finally, to complete the picture of all different embeddings of the
same four-dimensional solutions in the $N=8$ theory being related by
$U$~duality transformations (so that there is only one inquivalent embedding
for each solution) one should also study R-R and mixed embeddings. It
is, however, unlikely that the picture will change from what we have
presented above, since $U$~duality
treats both sectors on the same footing and also interchanges them.

\begin{table}
\begin{center}
\begin{tabular}{cc|c|c||}
\hline\hline
& & & \\
type~IIA${}^{1}$ & $\rightarrow$ & $\hat{\psi}_{\hat{a}}^{(+)1}$ &
$\hat{\psi}_{\hat{a}}^{(-)1}$ \\
\hline
& & & \\
type~IIA${}^{2}$ & $\rightarrow$ & $\hat{\psi}_{\hat{a}}^{(+)2}$ &
$\hat{\psi}_{\hat{a}}^{(-)2}$ \\
\hline
& & & \\
& & $\uparrow$ & $\uparrow$ \\
& & & \\
& & type~IIB${}^{(+)}$ & type~IIB${}^{(-)}$ \\
\end{tabular}
\end{center}

\caption[N=2 Supergravities sectors]
{Here we have represented symbolically the gravitini of four possible
$N=2,D=10$ theories that we can define in such a way that the ``common
sectors'' are in the correponding intersection of row and column. Each
common sector loosely corresponds to an $N=1,D=10$ theory.}
\label{tab-sectors}
\end{table}

%%%%%%%%%%%%%%%%%%%%%%%%%%%%%%%%%%%%%%%%%%%%%%%%%%%%%%%%%%%%%%%%%%%%%%

\section*{Acknowledgements}

We would like to thank Mike Duff, Joachim Rahmfeld and Paul Townsend for
helpful discussions.  One of us (T.O.) is indebted to M.M.~Fern\'andez for her
permanent support.  R.K.~would like to thank the East Carolina University
Department of Mathematics for their hospitality.
R.K.~was supported by a World Laboratory Fellowship.

%%%%%%%%%%%%%%%%%%%%%%%%%%%%%%%%%%%%%%%%%%%%%%%%%%%%%%%%%%%%%%%%%%%%%
%%%%%%%%%%%%%%%%%%%%%%%%%%%%%%%%%%%%%%%%%%%%%%%%%%%%%%%%%%%%%%%%%%%%%
%%%%%%%%%%%%%%%%%%%%%%%%%%%%%%%%%%%%%%%%%%%%%%%%%%%%%%%%%%%%%%%%%%%%%

\appendix

%%%%%%%%%%%%%%%%%%%%%%%%%%%%%%%%%%%%%%%%%%%%%%%%%%%%%%%%%%%%%%%%%%%%%
%%%%%%%%%%%%%%%%%%%%%%%%%%%%%%%%%%%%%%%%%%%%%%%%%%%%%%%%%%%%%%%%%%%%%
%%%%%%%%%%%%%%%%%%%%%%%%%%%%%%%%%%%%%%%%%%%%%%%%%%%%%%%%%%%%%%%%%%%%%

\section{Conventions}
\label{sec-conventions}

We denote with two hats eleven-dimensional objects, with one hat,
ten-dimensional objects and with no hats four-dimensional objects. Greek
and underlined latin or numerical indices are always world indices, and
simple latin or numerical indices are Lorentz indices. We reserve
the indices $i,j,k$ for the values $1,2,3$. The completely antisymmetric
symbol $\epsilon_{ijk}$ is defined by $\epsilon_{123}=+1$.

We use the mostly minuses signature $(+--\ldots-)$ and work in the
string-frame metric, except when we use the canonical metric that we
denote by a tilde $\tilde{g}_{\mu\nu}$.  The antisymmetric Levi-Civita
tensor ${\hat{\hat \epsilon}}$ is defined by

\begin{equation}
\hat{\hat{\epsilon}}^{\hat{\hat{\mu}}_{0}\ldots
\hat{\hat{\mu}}_{10}} = 1.
\end{equation}

Our spin connection $\omega$ (in $D$ dimensions) is defined by

\begin{equation}
\omega_{\mu}{}^{ab}(e)= -e^{\nu[a}
\left( \partial_{\mu}e_{\nu}{}^{b]} -\partial_{\nu}e_\mu{}^{b]}
\right) - e^{\rho[a} e^{\sigma b]}
\left( \partial_{\sigma} e_{c\rho} \right) e_{\mu}{}^{c}\ .
\end{equation}

\noindent The curvature tensor corresponding to this spin connection
field is defined by

\begin{equation}
R_{\mu\nu}{}^{ab}(\omega)=
2\partial_{[\mu}\omega_{\nu]}{}^{ab}-
2\omega_{[\mu}{}^{ac}
\omega_{\nu]c}{}^b\, ,
\hskip 1truecm
R(\omega)\equiv
e^{\mu}{}_{a}e^{\nu}{}_{b} R_{\mu\nu}{}^{ab}(\omega)\, .
\end{equation}

Indices not shown are assumed to be completely antisymmetrized
world indices. Thus, for instance,

\begin{equation}
\partial H = {\textstyle\frac{1}{2}}F^{(1)m}F^{(2)}{}_{m}\, ,
\end{equation}

\noindent (the Bianchi identity for $B$ Eq.~(\ref{eq:bianchis}) stands
for

\begin{equation}
\partial_{[\alpha} H_{\beta\gamma\delta]} =
{\textstyle\frac{1}{2}}F^{(1)m}{}_{[\alpha\beta}
F^{(2)}{}_{m\gamma\delta]}\, .
\end{equation}

%%%%%%%%%%%%%%%%%%%%%%%%%%%%%%%%%%%%%%%%%%%%%%%%%%%%%%%%%%%%%%%%%%%%%

\section{The type~IIA bosonic action and supersymmetry transformation
rules in the string frame, their truncation to $N=1$ and further
reduction to $D=4$}
\label{sec-IIAsusy}

%%%%%%%%%%%%%%%%%%%%%%%%%%%%%%%%%%%%%%%%%%%%%%%%%%%%%%%%%%%%%%%%%%%%%%%

\subsection{Dimensional reduction from $D=11$ to $D=10$}

The best way to obtain the supersymmetry transformation laws of the
ten-dimensional type~IIA theory in the string frame is by direct
dimensional reduction of $N=1,D=11$ supergravity, since we know that the
dilaton is just a function of the only scalar modulus field that appears
\cite{kn:W,kn:BHO}.

The bosonic fields of $N=1,D=11$ supergravity \cite{kn:CJS} are the
elfbein and a three-form potential

\begin{equation}
\left\{\hat{\hat{e}}_{\hat{\hat{\mu}}}{}^{\hat{\hat{a}}},
\hat{\hat{C}}_{\hat{\hat{\mu}}\hat{\hat{\nu}}\hat{\hat{\rho}}}
\right\}\, .
\end{equation}

The field strength of the three-form is

\begin{equation}
\hat{\hat{G}} =\partial \hat{\hat{C}}\, ,
\end{equation}

\noindent and the action for these bosonic fields is

\begin{equation}
\hat{\hat{S}}= {\textstyle\frac{1}{2}}\int d^{11}x\
\sqrt{\hat{\hat{g}}}\ \left[-\hat{\hat{R}} +a_{1}\hat{\hat{G}}^{2}
+a_{2}\frac{1}{\sqrt{\hat{\hat{g}}}}
\hat{\hat{\epsilon}} \hat{\hat{G}} \hat{\hat{G}} \hat{\hat{C}}
\right]\, .
\label{eq:11daction}
\end{equation}

\noindent where

\begin{equation}
a_{1}^{3}/a_{2}^{2}=2^{8}3^{5}\, , a_{1}>0\, .
\end{equation}

The only fermionic field of this theory is the gravitino
$\hat{\hat{\psi}}_{\hat{\hat{\mu}}}$, whose supersymmetry transformation
law, for purely bosonic configurations is\footnote{Our gamma matrices
are in a purely imaginary Majorana representation and have the
anticommutation relations $\left\{ \hat{\hat{\gamma}}^{\hat{\hat{a}}},
\hat{\hat{\gamma}}^{\hat{\hat{b}}}\right\} =
+2\hat{\hat{\eta}}^{\hat{\hat{a}}\hat{\hat{b}}}$.}

\begin{equation}
{\textstyle\frac{1}{\sqrt{2}}} \delta_{\epsilon}
\hat{\hat{\psi}}_{\hat{\hat{\mu}}}
= \hat{\hat{\nabla}}_{\hat{\hat{\mu}}}\hat{\hat{\epsilon}}
-6i\frac{a_{2}}{a_{1}}
\left(\hat{\hat{\gamma}}^{\hat{\hat{\alpha}} \hat{\hat{\beta}}
\hat{\hat{\gamma}} \hat{\hat{\delta}}}{}_{\hat{\hat{\mu}}}
-8 \hat{\hat{\gamma}}^{\hat{\hat{\beta}} \hat{\hat{\gamma}}
\hat{\hat{\delta}}}
\delta^{\hat{\hat{\alpha}}}{}_{\hat{\hat{\mu}}} \right)
\hat{\hat{G}}_{\hat{\hat{\alpha}} \hat{\hat{\beta}} \hat{\hat{\gamma}}
\hat{\hat{\delta}}}\ \hat{\hat{\epsilon}}\, ,
\end{equation}

The dimensional reduction has been explicitly performed in
Ref.~\cite{kn:BHO}. We can use the same ansatz for the elfbein to get
the same result for the $D=10$ type~IIA action in the string frame. We
refer the reader to that reference for details on the defininitions of
the ten-dimensional fields and field strengths. On top of that we make
the identifications

\begin{eqnarray}
\hat{\hat{\gamma}}^{\hat{a}} & = &
\hat{\gamma}^{\hat{a}}\, ,\,\,\, \hat{a}=0,\ldots,9\, ,
\\
& &
\nonumber \\
\hat{\hat{\gamma}}^{10} & = &
-i\ \hat{\gamma}_{11}= \hat{\gamma}^{0}\ldots\hat{\gamma}^{9}\, ,
\end{eqnarray}

\noindent so $\hat{\gamma}_{11}$ satisfies $(\hat{\gamma}_{11})^{2}=+1$
and can be used to define ten-dimensional chiralities. We define
the ten-dimensional spinors\footnote{Observe that these definitions
differ from those in Ref.~\cite{kn:HN} not only by powers of
$e^{\hat{\phi}}$ but also, in the gravitino case, by the relative
factor between $\hat{\hat{\psi}}_{\hat{a}}$ and
$\hat{\gamma}_{\hat{a}}\hat{\gamma}_{11}\hat{\hat{\psi}}_{11}$. Both
differences are caused by the fact that we are working
in the string frame.}

\begin{eqnarray}
\hat{\epsilon} & = &
e^{\frac{1}{6}\hat{\phi}}\ \hat{\hat{\epsilon}}\, ,
\\
& &
\nonumber \\
\hat{\psi}_{\hat{a}} & = &
{\textstyle\frac{1}{\sqrt{2}}} e^{-\frac{1}{6}\hat{\phi}}
\left(\hat{\hat{\psi}}_{\hat{a}} -{\textstyle\frac{i}{2}}
\hat{\gamma}_{\hat{a}} \hat{\gamma}_{11}\hat{\hat{\psi}}_{11}
\right)\, ,
\\
& &
\nonumber \\
\hat{\lambda} & = &
{\textstyle\frac{-3i}{\sqrt{2}}} e^{-\frac{1}{6}\hat{\phi}}
\hat{\gamma}_{11} \hat{\hat{\psi}}_{11}\, ,
\end{eqnarray}

\noindent set

\begin{equation}
a_{1}=\frac{3}{4}\,\,\,\, \Rightarrow a_{2}=2^{-7}3^{-1}\, ,
\end{equation}

\noindent and get the action \cite{kn:BHO}

\begin{eqnarray}
\hat{S}^{IIA} & = &
{\textstyle\frac{1}{2}} \int d^{10}x\
\sqrt{|\hat{g}|} \left\{ e^{-2\hat{\phi}}
\left[ -\hat{R} +4\left( \partial\hat{\phi} \right)^{2}
-{\textstyle\frac{3}{4}} \hat{H}^{2}\right]\right.
\nonumber \\
& &
\nonumber \\
& &
+{\textstyle\frac{1}{4}} \hat{F}^{2}
+{\textstyle\frac{3}{4}}\hat{G}^{2}
+{\textstyle\frac{1}{64}} \frac{\hat{\epsilon}}{\sqrt{-\hat{g}}}\
\partial\hat{C}\partial\hat{C}\hat{B}\biggr \}\, ,
\label{eq:iiaaction}
\end{eqnarray}

\noindent and supersymmetry transformation laws of the gravitino
$\hat{\psi}_{\hat{a}}$ and dilatino $\hat{\lambda}$ fields

\begin{eqnarray}
\delta_{\epsilon} \hat{\psi}_{\hat{a}} & = &
\partial_{\hat{a}}\hat{\epsilon} -{\textstyle\frac{1}{4}}
\left( \hat{\omega}_{\hat{a}\hat{b}\hat{c}} -{\textstyle\frac{3}{2}}
\hat{H}_{\hat{a}\hat{b}\hat{c}} \hat{\gamma}_{11}\right)
\hat{\gamma}^{\hat{b}\hat{c}}\ \hat{\epsilon}
\\
& &
\nonumber \\
& &
-{\textstyle\frac{i}{16}}e^{\hat{\phi}}
\left(\hat{\gamma}_{\hat{a}}{}^{\hat{b}\hat{c}}
-2\delta_{\hat{a}}{}^{\hat{b}}\hat{\gamma}^{\hat{c}}
\right)\hat{\gamma}_{11}\hat{F}_{\hat{b}\hat{c}}\ \hat{\epsilon}
\\
& &
\nonumber \\
& &
-{\textstyle\frac{i}{32}}e^{\hat{\phi}}
\left(\hat{\gamma}_{\hat{a}}{}^{\hat{b}\hat{c}\hat{d}\hat{e}}
-4\delta_{\hat{a}}{}^{\hat{b}}\hat{\gamma}^{\hat{c}\hat{e}}
\right)\hat{G}_{\hat{b}\hat{c}\hat{d}\hat{e}}\ \hat{\epsilon}\, ,
\\
& &
\\
\delta_{\epsilon} \hat{\lambda} & = &
\left(\not\!\partial\hat{\phi}
-{\textstyle\frac{1}{4}}\not\!\!\hat{H}\hat{\gamma}_{11}
\right)\hat{\epsilon}
-{\textstyle\frac{3i}{8}}e^{\hat{\phi}}
\left(\not\!\!\hat{F}\hat{\gamma}_{11}
+{\textstyle\frac{1}{6}}\not\!\!\hat{G}\right)\hat{\epsilon}\, .
\end{eqnarray}

%%%%%%%%%%%%%%%%%%%%%%%%%%%%%%%%%%%%%%%%%%%%%%%%%%%%%%%%%%%%%%%%%%%%%%%

\subsection{Truncation to the $N=1$ theory}

When all RR fields ($\hat{C}_{\hat{\mu} \hat{\nu} \hat{\rho}},
\hat{A}_{\hat{\mu}}$) are set to zero (which is a consistent
truncation), the bosonic action of the type~IIA theory
Eq.~(\ref{eq:iiaaction}) reduces to that of the $N=1$ theory, which only
contains the NS-NS fields $\hat{g}_{\hat{\mu}\hat{\nu}},
\hat{B}_{\hat{\mu}\hat{\nu}}, \hat{\phi}$:

\begin{equation}
\hat{S}^{N=1}  =
{\textstyle\frac{1}{2}} \int d^{10}x\
\sqrt{|\hat{g}|}\ e^{-2\hat{\phi}}
\left[ -\hat{R} +4\left( \partial\hat{\phi} \right)^{2}
-{\textstyle\frac{3}{4}} \hat{H}^{2}\right]
\label{eq:n=1action}
\end{equation}

\noindent In the supersymmetry transformation rules, though, both
chiralities are still present.  If we split all spinors in their
positive ${}^{(+)}$ and negative ${}^{(-)}$ chirality halves

\begin{equation}
\hat{\epsilon}=\hat{\epsilon}^{(+)}+\hat{\epsilon}^{(-)}\, ,
\hspace{.5cm}
\hat{\gamma}_{11}\ \hat{\epsilon}^{(\pm)} =
\pm \hat{\epsilon}^{(\pm)}\, ,
\end{equation}

\noindent etc.~we get

\begin{eqnarray}
\delta_{\epsilon} \hat{\psi}^{(\pm)}_{\hat{a}} & = &
\hat{\nabla}_{\hat{a}}^{(\pm)} \hat{\epsilon}^{(\pm)}\, ,
\nonumber \\
& &
\nonumber \\
\delta_{\epsilon} \hat{\lambda}^{(\pm)} & = &
\left(\not\!\partial\hat{\phi}
\pm {\textstyle\frac{1}{4}}\not\!\!\hat{H}\right)
\hat{\epsilon}^{(\pm)}\, ,
\label{eq:norrrules}
\end{eqnarray}

\noindent where $\hat{\nabla}_{\hat{a}}^{(\pm)}$ are the covariant
derivatives associated to the two torsionful spin connections

\begin{equation}
\hat{\Omega}^{(\pm)}_{\hat{a} \hat{b} \hat{c}}
=\hat{\omega}_{\hat{a} \hat{b} \hat{c}}
\mp{\textstyle\frac{3}{2}} \hat{H}_{\hat{a} \hat{b} \hat{c}}\, .
\end{equation}

Eqs.~(\ref{eq:norrrules}) are just the supersymmetry transformation
rules of the gravitino and dilatino of two $N=1$ theories of different
chiralities.  Both are related by a change of sign of the axion and a
change in the chirality of the supersymmetry parameter $\hat{\epsilon}$.
This transformation is a duality symmetry of the $N=2A$ theory and a
string/string duality symmetry between two different $N=1$ theories of
opposite chirality in its own right\cite{kn:BJO}: $C$~duality.

%%%%%%%%%%%%%%%%%%%%%%%%%%%%%%%%%%%%%%%%%%%%%%%%%%%%%%%%%%%%%%%%%%%%%%%

\subsection{Further reduction from $D=10$ to $D=4$}

The dimensional reduction of $N=1,D=10$ supergravity to $10-d$
dimensions in the string frame was performed in Ref.~\cite{kn:MS} (in
the canonical frame it was done in Ref.~\cite{kn:C}).  Here we just
quote their result for the four-dimensional action using our
conventions, and give the relations between ten- and four-dimensional
fields that allow us to uplift four-dimensional solutions to ten
dimensions.

The four-dimensional action is

\begin{eqnarray}
S & = & {\textstyle\frac{1}{2}} \int d^{4}x\ \sqrt{|g|}\ e^{-2\phi}
\left\{ -R +4(\partial\phi)^{2} -{\textstyle\frac{3}{4}}H^{2} \right.
\nonumber \\
& &
\nonumber \\
& &
+{\textstyle\frac{1}{4}}\left[\partial G_{mn} \partial G^{mn} -
G^{mn} G^{pq} \partial B_{mp} \partial B_{nq}\right]
\nonumber \\
& &
\nonumber \\
& &
\left.
-{\textstyle\frac{1}{4}} \left[ G_{mn}F^{(1)m}F^{(1)n}
+G^{mn}{\cal F}_{m}{\cal F}_{m} \right]
\right\}\, ,
\label{eq:4daction}
\end{eqnarray}

\noindent where the vector and axion field-strenghts are

\begin{equation}
\begin{array}{rclrcl}
F^{(1)m}
&
=
&
2\partial A^{(1)m}\, ,
&
H
&
=
&
\partial B -{\textstyle\frac{1}{2}}A^{(1)m}F^{(2)}{}_{m}
-{\textstyle\frac{1}{2}}A^{(2)}{}_{m}F^{(1)m}\, ,
\\
& & & & & \\
F^{(2)}{}_{m}
&
=
&
2\partial A^{(2)}{}_{m}\, ,
&
{\cal F}_{m}
&
=
&
F^{(2)}{}_{m} + F^{(1)n}B_{nm}\, ,
\\
\end{array}
\end{equation}

If we are given the four-dimensional fields $g_{\mu\nu}, B_{\mu\nu},
A^{(1)m}{}_{\mu}, A^{(2)}{}_{m\mu}, G_{mn}, B_{mn}$ and $\phi$ of a
solution, the ten-dimensional fields of the corresponding
ten-dimensional solutions can be found by using

\begin{equation}
\begin{array}{rclrcl}
\hat{g}_{\mu\nu}
&
=
&
g_{\mu\nu} + A^{(1)m}{}_{\mu}A^{(1)n}{}_{\nu}G_{mn}\, ,
&
\hat{g}_{mn}
&
=
&
G_{mn}\, ,
\\
& & & & & \\
\hat{B}_{\mu\nu}
&
=
&
B_{\mu\nu} +A^{(1)m}{}_{\mu}A^{(1)n}{}_{\nu}B_{mn}
-A^{(1)m}{}_{\mu}A^{(2)}{}_{m\nu}\, ,
&
\hat{B}_{mn}
&
=
&
B_{mn}
\\
& & & & & \\
\hat{B}_{\mu m}
&
=
&
A^{(2)}{}_{m\mu} + A^{(1)n}{}_{\mu}B_{nm}\, ,
&
\hat{g}_{\mu m}
&
=
&
A^{(1)n}{}_{\mu}G_{nm}\, ,
\\
& & & & & \\
\hat{\phi}
&
=
&
\phi +{\textstyle\frac{1}{4}}\log{{\rm det} |G|}\, .
& & & \\
\end{array}
\label{eq:uplifting}
\end{equation}

It is also useful to perform the dimensional reduction of the $N=1,D=10$
supersymmetry transformation rules to identify which are the six vector
fields that belong to the $N=4,D=4$ gravity supermultiplet and the six
vector fields that belong to the six additional $N=4,D=4$ vector
supermultiplets.  For this purpose, it is not necessary to reduce the
spinor indices and thus we will keep the ten-dimensional gamma matrices
and spinor indices.  It is also sufficient to reduce the positive
chirality theory (the rules of the theory with opposite ten-dimensional
chirality can be obtained by a change of sign of $B_{\mu\nu}, B_{mn}$
and $A^{(2)}_{\mu m}$). The gravitini, dilatini and photini
supersymmetry transformation rules are, respectively

\begin{eqnarray}
\delta_{\epsilon} \hat{\psi}_{a}^{(+)} & = &
\nabla_{a}^{(+)}\hat{\epsilon}^{(+)}
-{\textstyle\frac{1}{4}}
\left(\not\!\! F^{1}_{m} -\not\!\!{\cal F}_{m} \right)
e_{i}^{m} \hat{\gamma}^{i}\ \hat{\epsilon}^{(+)}
\nonumber \\
& &
\nonumber \\
& &
+\left(\partial_{a}e_{mj} +\partial_{a}B_{mn}e_{j}^{n}\right) e_{i}^{m}
\hat{\gamma}^{ij}\ \hat{\epsilon}^{(+)}\, ,
\nonumber \\
& &
\nonumber \\
\delta_{\epsilon} \left( \hat{\lambda}^{(+)}
-\hat{\gamma}^{i} \hat{\psi}_{i}^{(+)} \right) & = &
\left\{\not\!\partial\phi
+{\textstyle\frac{1}{4}}\not\!\! H
-{\textstyle\frac{1}{8}}\left(\not\!\! F^{(1)}_{m}
-\not\!\!{\cal F}_{m}\right) e_{i}^{m}\hat{\gamma}^{i}
\right\}\hat{\epsilon}^{(+)}\, ,
\\
& &
\nonumber \\
\delta_{\epsilon} \hat{\psi}_{i}^{(+)} & = &
{\textstyle\frac{1}{8}}\left(\not\!\! F^{(1)}_{m}
+\not\!\!{\cal F}_{m}\right)
e_{i}^{m}\hat{\gamma}^{i}\ \hat{\epsilon}^{(+)}
\nonumber \\
& &
\nonumber \\
& &
-{\textstyle\frac{1}{4}}
\left( \not\!\partial G_{mn} +\not\!\partial B_{mn} \right)
e_{i}^{m}e_{j}^{n}\hat{\gamma}^{j}\ \hat{\epsilon}^{(+)}\, .
\end{eqnarray}

This means that the supergravity vector fields are in the combinations

\begin{equation}
F^{(1)m}-{\cal F}_{m}\, ,
\end{equation}

\noindent and the matter vector fields are in the combinations

\begin{equation}
F^{(1)m}+{\cal F}_{m}\, .
\end{equation}

These combinations interchange their roles under $C$~duality.

%%%%%%%%%%%%%%%%%%%%%%%%%%%%%%%%%%%%%%%%%%%%%%%%%%%%%%%%%%%%%%%%%%%%%%%

\section{Spin connection coefficients}
\label{sec-spincon}

Most of the ten-dimensional metrics we have met are of the general form

\begin{eqnarray}
d\hat{s}^{2} & = & V^{d}dt^{2} -W^{b}d\vec{x}^{2}
-\left(V^{-\frac{d}{2}} dx^{\underline{4}}
+n\ V^{\frac{d}{2}}dt\right)^{2}
\nonumber \\
& &
\nonumber \\
& &
-W^{c}\left(dx^{\underline{5}}
+m\ W_{\underline{i}}dx^{\underline{i}}\right)^{2}
-dx^{\underline{I}}dx^{\underline{I}}\, ,
\end{eqnarray}

\noindent with $I=6,\ldots,9$. We ignore these directions since
they are flat. Choosing the zehnbein one-form basis

\begin{equation}
\begin{array}{rclrcl}
\hat{e}^{0}
&
=
&
V^{\frac{d}{2}}dt\, ,
&
\hat{e}^{4}
&
=
&
V^{-\frac{d}{2}}dx^{\underline{4}} +n\ V^{\frac{d}{2}}dt\, ,
\\
& & & & & \\
\hat{e}^{i}
&
=
&
W^{\frac{b}{2}}dx^{\underline{i}}\, ,
&
\hat{e}^{5}
&
=
&
W^{\frac{c}{2}}\left(dx^{\underline{5}} +m\
W_{\underline{i}}\right)\, ,
\end{array}
\end{equation}

\noindent we get the following non-vanishing components of the spin
connection one-form

\begin{eqnarray}
\hat{\omega}^{0i} & = & {\textstyle\frac{d}{2}}
W^{-\frac{b}{2}}V^{-1}\partial_{\underline{i}}V
\left(\hat{e}^{0} -n\ \hat{e}^{4}\right)\, ,
\\
& &
\nonumber \\
\hat{\omega}^{04} & = & -{\textstyle\frac{nd}{2}}
W^{-\frac{b}{2}}V^{-1}\partial_{\underline{i}}V\
\hat{e}^{i}\, ,
\\
& &
\nonumber \\
\hat{\omega}^{ij} & = & +bW^{-\frac{b}{2}-1}
\delta_{k[i}\partial_{\underline{j}]}W\ \hat{e}^{k} -m\
W^{\frac{c}{2}-b}
\partial_{[\underline{i}} W_{\underline{j}]} \hat{e}^{5}\, ,
\\
& &
\nonumber \\
\hat{\omega}^{i4} & = & -{\textstyle\frac{nd}{2}}
W^{-\frac{b}{2}} V^{-1} \partial_{\underline{i}}V
\left(\hat{e}^{0}-n\ \hat{e}^{4}\right)\, ,
\\
& &
\nonumber \\
\hat{\omega}^{i5} & = & -m\ W^{\frac{c}{2}-b}
\partial_{[\underline{i}} W_{\underline{j}]} \hat{e}^{j}
-{\textstyle\frac{c}{2}} W^{-\frac{b}{2}-1}
\partial_{\underline{i}}W\hat{e}^{5}\, .
\end{eqnarray}

%%%%%%%%%%%%%%%%%%%%%%%%%%%%%%%%%%%%%%%%%%%%%%%%%%%%%%%%%%%%%%%%%%%%%%%


\begin{thebibliography}{30}

\bibitem{kn:DK}  M.J.~Duff and R.R.~Khuri, {\sl Four-Dimensional
                 String/String Duality}, {\it Nucl. Phys.}~{\bf B411}
                 (1994) 473.

\bibitem{kn:PR}  M.J.~Duff, R.R.~Khuri and J.X.~Lu, {\sl String
                 Solitons}, {\it Phys.~Rep.}~{\bf 4\& 5} (1995) 213.

\bibitem{kn:HT} C.M.~Hull and P.K.~Townsend, {\sl Unity of Superstring
                 Dualities}, {\it Nucl.~Phys.}~{\bf B438} (1995) 109.

\bibitem{kn:W} E.~Witten, {\sl String Dynamics in various dimensions},
               {\it Nucl.~Phys.}~{\bf B443} (1995) 85.

\bibitem{kn:BBKOBSHAAGB} I.~Bakas, {\sl Space-Time Interpretations of
                        $S$~Duality and Supersymmetry Violations of
                        $T$~Duality}, {\it Phys.~Lett.}~{\bf 343B}
                        (1995) 103-112.\\
                        E.~Bergshoeff, R.~Kallosh and T.~Ort\'{\i}n,
                        {\sl Duality Versus Supersymmetry and
                        Compactification}, {\it Phys.~Rev.}~{\bf D51}
                        (1995) 3009.\\
                        I.~Bakas and K.~Sfetsos, {\sl $T$~Duality and
                        World-Sheet Supersymmetry}, {\it
                        Phys.~Lett.}~{\bf 349B} (1995) 448-457.\\
                        S.F.~Hassan, {\sl T-duality And Non-Local
                        Supersymmetry}, Report CERN-T/95-98 and
                        {\tt hep-th/9504148}.\\
                        E.~\'Alvarez, L.~\'Alvarez-Gaum\'e and I.~Bakas,
                        {\sl $T$~Duality and Space-Time Supersymmetry},
                        Report CERN-TH/95-258,  FTUAM/95-34 and {\tt
                        hep-th/9510028}.

\bibitem{kn:O1} T.~Ort\'{\i}n, {\sl Electric-Magnetic Duality  and
                Supersymmetry in Stringy Black Holes}, {\it
                Phys.~Rev.}~{\bf D47} (1993) 3136-3143.

\bibitem{kn:O2} T.~Ort\'{\i}n, {\sl SL(2,\R)--Duality Covariance of
                Killing Spinors in Axion--Dilaton Black Holes}, {\it
                Physical Review} {\bf D51}, (1995) 790--794.

\bibitem{kn:IPT} J.M.~Izquierdo, N.D.~Lambert, G.~Papadopoulos and
                 P.K.~Townsend, {\sl Dyonic membranes}, Report
                 DAMTP-R-95-40 and {\tt hep-th/9508177}.

\bibitem{kn:SS}  M.J.~Duff, S.~Ferrara, R.R.~Khuri and J.~Rahmfeld,
                 {\sl Supersymmetry and Dual String Solitons},
                 {\it Phys.~Lett.} ~{\bf 356B} (1995) 479.

\bibitem{kn:GWG} G.W.~Gibbons, {\sl Aspects of Supergravity Theories},
                 (three lectures) in: {\sl Supersymmetry, Supergravity
                 and Related Topics},  eds.~F.~del Aguila, J.~de
                 Azc\'arraga and L.~Ib\'a\~nez, World Scientific,
                 Singapore, 1985, page 147.

\bibitem{kn:KLOPP} R.~Kallosh, A.~Linde, T.~Ort\'{\i}n, A.~Peet and
                   A.~Van Proeyen, {\sl Supersymmetry as a Cosmic
                   Censor}, {\it Phys.~Rev.} {\bf D46} (1992) 5278-5302.

\bibitem{kn:DI}  R.R.~Khuri, {\sl Remark on String Solitons},
                 {\it Phys. Rev.}~{\bf D48} (1993) 2947.


\bibitem{kn:GHS} D.~Garfinkle, G.~Horowitz and A.~Strominger, {\sl
                 Charged Black Holes in String Theory},
                 {\it Phys.~Rev.}~{\bf D43} (1991) 3140, {\bf Erratum},
                 {\it ibid.}~{\bf D45} (1992) 3888.

\bibitem{kn:KO} R.R.~Khuri and T.~Ort\'{\i}n, {\sl A Non-Supersymmetric
                Dyonic Extreme Reissner-Nordstr\"om Black Hole}, Report
                CERN-TH/95-347 to appear.

\bibitem{kn:BJO} E.~Bergshoeff, B.~Janssen and T.~Ort\'{\i}n, {\sl
                 Solution-Generating Transformations and the String
                 Effective Action}, {\tt hep-th/9506156} (to be
                 published in {\it Phys.~Rev.}~{\bf D15}).

\bibitem{kn:CYKK} M.~Cveti\u{c} and D.~Youm, {\sl Static Four-Dimensonal
                  Abelian Black Holes in Kaluza-Klein Theory}, 
                  {\it Phys.~Rev.}~{\bf D52} (1995) 2144-2149.\\
                  M.~Cveti\u{c} and D.~Youm, {\sl Kaluza-Klein Black Holes
                  within Heterotic String Theory on a Torus}, 
                  {\it Phys.~Rev.}~{\bf D52} (1995) 2574-2576.\\
                  M.~Cveti\u{c} and D.~Youm, {\sl All the Four-Dimensonal
                  Static Spherically Symmetric Solutions of 
                  Abelian Kaluza-Klein Theory}, 
                  {\it Phys.~Rev. Lett.}~{\bf 75} (1995) 4165-4168.
                  
\bibitem{kn:CYN8} M.~Cveti\u{c} and D.~Youm, {\sl Four-Dimensional
                  Supersymmetric Dyonic Black Holes in Eleven-Dimensional
                  Supergravity}, {\it Nucl.~Phys.}~{\bf B453} (1995)
                  259-280.\\
                  M.~Cveti\u{c} and D.~Youm, {\sl BPS Saturated Dyonic
                  Black Holes of $N=8$ Supergravity Vacua}, talk given at
                  the SUSY'95 Conference, May 8-12 1995 in Paris
                  (France) and the Conference on $S$~Duality and Mirror
                  Symmetry, May 22-26 1995 in Trieste (Italy), Report
                  IASSNS-HEP-95/80, PUPT-1566 and {\tt hep-th/9510098}.

\bibitem{kn:CY4P} M.~Cveti\u{c} and D.~Youm, {\sl Dyonic BPS Saturated
                  Black Holes of Heterotic String on a Six Torus}, Report
                  UPR-0672-TA and {\tt hep-th/9507090} (to be
                  published in {\it Phys.~Rev.}~{\bf D15}).\\
                  M.~Cveti\u{c} and A.~A.~Tseytlin, {\sl General Class of
                  BPS Saturated Dyonic Black Holes as Exact Superstring 
                  Solutions}, Report IASSNS-HEP-95/79, Imperial/TP/95-96/4
                  and {\tt hep-th/9510097} (to be published in 
                  {\it Phys.~Lett.}~{\bf B}).
      

\bibitem{kn:CY5P} M.~Cveti\u{c} and A.~A.~Tseytlin, {\sl Solitonic Strings
                  and BPS Saturated Dyonic Black Holes}, 
                  Report IASSNS-HEP-95/96, Imperial/TP/95-96/14
                  and {\tt hep-th/9512031}.\\
                  M.~Cveti\u{c} and D.~Youm, {\sl All the Statis Spherically
                  Symmetric Black Holes of Heterotic String on a Six Torus},
                  Report IASSNS-HEP-95/107 and {\tt hep-th/9512127}.
                  
\bibitem{kn:CYSS} M.~Cveti\u{c} and D.~Youm, {\sl Singular BPS Saturated States
                  and Enhanced Symmetries of Four-Dimensional N=4                         
                  Supersymmetric String Vacua}, {\it Phys.~Lett.}~{\bf 359B}
                  (1995) 87-92.\\
                  M.~Cveti\u{c} and D.~Youm, {\sl BPS Saturated and Non-Extreme
                  States in Abelian Kaluza-Klein Theory and Effective N=4
                  Supersymmetric String Vacua}, talk given at
                  the Strings'95 Conference, March 13-18 1995 in Los Angeles,                                                
                  CA (USA), Report UPR-675-T, NSF-ITP-95-74
                  and {\tt hep-th/9508058}.


\bibitem{kn:HW} C.F.~Holzhey and F.~Wilczek, {\sl Black Holes as
                Elementary Particles}, {\it Nucl.~Phys.}~{\bf B380}
                (1992) 447-477.

\bibitem{kn:S1} K.~Shiraishi, {\sl Multicentered Solution for Maximally
                Charged Dilaton Black Holes in arbitrary Dimensions},
                {\it J.~Math.~Phys.}~{\bf 34}(4) (1993) 1480.

\bibitem{kn:GGM} G.W.~Gibbons, {\sl Antigravitating Black Hole Solitons
                 with Scalar Hair in $N=4$ Supergravity}, {\it
                 Nucl.~Phys.}~{\bf B207}, (1982) 337.\\
                 G.W.~Gibbons and K.~Maeda, {\sl Black Holes and
                 Membranes in Higher Dimensional Theories with Dilaton
                 Fields}, {\it Nucl.~Phys.}~{\bf B298}, (1988) 741.

\bibitem{kn:GKLTT} G.W.~Gibbons, D.~Kastor, L.A.J.~London, P.K.~Townsend
                   and J.~Traschen, {\sl Supersymmetric Selfgravitating
                   Solitons}, {\it Nucl.~Phys.}~{\bf B416} (1994)
                   850-880.

\bibitem{kn:HM}  R.R.~Khuri, {\sl Some Instanton Solutions in String
                 Theory}, {\it Phys.~Lett.} ~{\bf 259B} (1991) 261.\\
                 R.R.~Khuri, {\sl A Heterotic Multimonopole Solution
                 in String Theory}, {\it Nucl.~Phys.}~{\bf B387} (1992)
                 315.\\
                 M.J.~Duff, R.R.~Khuri, R.~Minasian and J.~Rahmfeld,
                 {\sl New Black Hole, String and Membrane Solutions
                  of the Four-Dimensional Heterotic String},
                 {\it Nucl.~Phys.} ~{\bf B418} (1994) 195.

\bibitem{kn:K}  R.R.~Khuri, {\sl Scattering of String Monopoles},
                {\it Phys.~Lett.}~{\bf 294B} (1992) 331.

\bibitem{kn:S2} K.~Shiraishi, {\sl Moduli Space Metric for Maximally
               Charged Dilaton Black Holes}, {\it Nucl.~Phys.}~{\bf 402}
               (1993) 399-410.

\bibitem{kn:G} J.~Gauntlett, {\sl Low-Energy Dynamics of
               Supersymmetric Solitons}, {\it Nucl.~Phys.}~{\bf B400}
               (1993) 103-125.

\bibitem{kn:DR} M.~Duff and J.~Rahmfeld, {\sl Massive String States as
                Extreme Black Holes}, {\it Phys.~Lett.}~{\bf 345B}
                (1995) 441.

\bibitem{kn:SCH} J.H.~Schwarz and A.~Sen, {\sl Duality Symmetries of
                 4-d Heterotic Strings}, {\it Phys.~Lett.}~{\bf 312B}
                 (1993) 105.\\
                 A.~Sen, {\sl Strong-Weak Coupling Duality in
                 Four-Dimensional String Theory}, {\it
                 Int.~J.~Mod.~Phys.}~{\bf A9} (1994) 3707.

\bibitem{kn:KM}  R.R.~Khuri and R.C.~Myers,  {\sl Dynamics of Extreme
                 Black Holes and Massive String States}, Report
                 McGill/95-38, CERN-TH/95-213 and {\tt hep-th/9508045}
                 (to be published in {\it Physical Review} {\bf D15}).\\
                 C.G.~Callan, J.M.~Maldacena and A.W.~Peet,
                 {\sl Extremal Black Holes as Fundamental Strings},
                 Report PUPT-1565 and {\tt hep-th/9510134}.\\
                 G.~Mandal and S.R.~Wadia, {\sl Black Hole Geometry
                 Around an Elementary BPS String State}, Report
                 TIFR-TH-95/61 and {\tt hep-th/9511218}.

\bibitem{kn:LIU} M.J.~Duff, J.T.~Liu and J.~Rahmfeld, {\sl
                 Four-Dimensional String/String/String Triality},
                 Report CTP-TAMU-27-95 and {\tt hep-th/9508094}.

\bibitem{kn:R} J.~Rahmfeld, {\sl Extremal Black Holes as Bound States},
               Report CTP-TAMU-51/95 and {\tt hep-th/9512089}.

\bibitem{kn:GGK} D.V.~Gal'tsov, A.A.~Garc\'{\i}a and O.V.~Kechkin, {\sl
                 Symmetries of the Stationary Einstein-Maxwell-dilaton
                 Theory}, {\it J.~Math.~Phys.}~{\bf 36} (1995)
                 5023-5041.

\bibitem{kn:BHO} E.~Bergshoeff, C.M.~Hull and T.~Ort\'{\i}n, {\sl
                 Duality in the Type~II Superstring Effective Action},
                 {\it Nucl.~Phys.} {\bf B451} (1995) 547-578.

\bibitem{kn:C} A.~Chamseddine, {\sl $N=4$ Supergravity Coupled to $N=4$
               Matter}, {\it Nucl.~Phys.}~{\bf B185}, (1981) 403.

\bibitem{kn:B} T.H.~Buscher, {\sl Quantum Corrections and Extended
               Supersymmetry in New Sigma Models}, {\it
               Phys.~Lett.}~{\bf 159B} (1985) 127.\\
               T.H.~Buscher, {\sl A Symmetry of the String Background
               Equations}, {\it Phys.~Lett.}~{\bf 194B} (1987) 59.\\
               T.H.~Buscher, {\sl Path Integral Derivation of Quantum
               Duality in Nonlinear Sigma Models}, {\it
               Phys.~Lett.}~{\bf 201B} (1988) 466.

\bibitem{kn:MS} J.~Maharana and J.H.~Schwarz, {\sl Non-Compact
                Symmetries in String Theory}, {\it Nucl.~Phys.}~{\bf
                B390} (1993) 3.

\bibitem{kn:BEK} E.~Bergshoeff, I.~Entrop and R.~Kallosh, {\sl Exact
                 Duality in the String Effective Action}, {\it
                 Phys.~Rev.}~{\bf D49} (1994) 6663.

\bibitem{kn:KM2} R.R.~Khuri and R.C.~Myers, {\sl Rusty Scatter Branes},
                 Report McGill/95-61, CERN-TH/95-331 and {\tt
                 hep-th/9512061}.

\bibitem{kn:LPSS} H.~Lu, C.N.~Pope, E.~Sezgin and K.S.~Stelle,
                  {\sl Stainless Super p-branes}, Report CTP-TAMU-31/95,
                  Imperial/TP/94-95/56, SISSA 97/95/EP and
                  and {\tt hep-th/9508042}.\\
                  M.J.~Duff, H.~Lu, C.N.~Pope and E.~Sezgin,
                  {\sl Supermembranes with Fewer Supersymmetries},
                  Report CTP-TAMU-35/95 and {\tt hep-th/9511162}.\\
                  H.~Lu, C.N.~Pope, E.~Sezgin and K.S.~Stelle,
                  {\sl Dilatonic p-brane Solitons}, Report CTP-TAMU-40/95,
                  Imperial/TP/95--96 and {\tt hep-th/9511203}.\\
                  H.~Lu and C.N.~Pope, {\sl p-brane Solitons in Maximal
                  Supergravities}, Report CTP-TAMU-47/95,
                  and {\tt hep-th/9512012}.

\bibitem{kn:BBO} E.~Bergshoeff, H.-J.~Boonstra and T.~Ort\'{\i}n, {\sl
                 $S$~Duality and Dyonic $p$-brane solutions in Type~II
                 String Theory}, {\tt hep-th/9508091} (to be published
                 in {\it Physical Review} {\bf D15}).

\bibitem{kn:CJS} E.~Cremmer, B.~Julia and J.~Scherk, {\sl Supergravity
                 in 11 Dimensions}, Phys.~Lett.~{\bf 76B} (1978) 409.

\bibitem{kn:HN} M.~Huq and M.A.~Namazie, {\sl Kaluza-Klein Supergravity
                in Ten Dimensions}, {\it Class.~Quantum Grav.}~{\bf 2}
                (1985) 293-308 and {\bf Corrigendum}, {\it
                Class.~Quantum Grav.}~{\bf 2} (1985) 597.

\end{thebibliography}
\end{document}